\documentclass[fleqn,usenatbib]{mnras}

\usepackage[T1]{fontenc}

\DeclareRobustCommand{\VAN}[3]{#2}
\let\VANthebibliography\thebibliography
\def\thebibliography{\DeclareRobustCommand{\VAN}[3]{##3}\VANthebibliography}

\usepackage{graphicx}	% Including figure files
\usepackage{amsmath}	% Advanced maths commands
\usepackage{amssymb}	% Extra maths symbols
\usepackage[dvipsnames]{xcolor}
\usepackage{newtxtext,newtxmath}

\title[Hot Jupiters in dense clusters]{Hot Jupiter formation in dense clusters: secular chaos in multi-planetary systems}

\author[Wang et al.]{
Yi-Han Wang,$^{1}$\thanks{E-mail: yihan.wang.1@stonybrook.edu}
Rosalba Perna$^{1,2}$
Nathan  W. C. Leigh,$^{3,4}$
Michael M. Shara,$^{4}$
\\
$^{1}$Department of Physics and Astronomy, Stony Brook University, Stony Brook, NY, 11794, USA\\
$^{2}$Center for Computational Astrophysics, Flatiron Institute, 162 5th Avenue, New York, NY 10010, USA\\
$^{3}$Departamento de Astronom\'ia, Facultad Ciencias F\'isicas y Matem\'aticas, Universidad de Concepci\'on, Av. Esteban Iturra s/n Barrio Universitario,\\ Casilla 160-C, Concepci\'on, Chile\\
$^{4}$Department of Astrophysics, American Museum of Natural History, Central Park West and 79th Street, New York, NY 10024\\
}

\date{Accepted XXX. Received YYY; in original form ZZZ}

\pubyear{2015}

\begin{document}
\label{firstpage}
\pagerange{\pageref{firstpage}--\pageref{lastpage}}
\maketitle

% Abstract of the paper
\begin{abstract}
Exoplanetary observations reveal that the occurrence rate of hot Jupiters is  correlated with star clustering. In star clusters, interactions between planetary systems and close fly-by stars can significantly change the architecture of primordially coplanar, circular planetary systems. Flybys in dense clusters have a significant impact on hot Jupiter formation via activation of high eccentricity excitation mechanisms such as the Zeipel-Lidov-Kozai (ZLK) effect and planet-planet scattering. Previous studies have shown that if there are two giant planets in the planetary system, close flybys, especially the ones with incidence angles with the planetary orbits close to 90 degrees, 
can efficiently activate the ZLK mechanism, thus triggering high eccentricity tidal migration and ultimately form hot Jupiters in star clusters. Here we extend our previous study with a multi-planet (triple) system. We perform high precision, high-accuracy few-body simulations of stellar flybys and subsequent planetary migration within the perturbed planetary systems using the code {\tt SpaceHub}. Our simulations demonstrate that a single close flyby on a multi-planet system in a cluster can activate secular chaos and ultimately lead to hot Jupiter formation via
high eccentricity migration. 
We find that the hot Jupiter formation rate per system increases with both the size of the planetary system as well as with the mass of the outer planet, and we quantify the relative formation fractions for a range of parameters. Hot Jupiters formed via secular chaos are expected to be accompanied by massive companions with very long periods.
Our study further shows that this flyby-induced secular chaos is preferred in low-density clusters where multi-planet systems are more likely to survive, and that it contributes a significant fraction of the hot Jupiter formation in star clusters compared to the flyby-induced ZLK mechanism.
\end{abstract}


%%%%%%%%%%%%%%%%%%%%%%%%%%%%%%%%%%%%%%%%%%%%%%%%%%

%%%%%%%%%%%%%%%%% BODY OF PAPER %%%%%%%%%%%%%%%%%%

\section{Introduction}

The discovery of giant gas exoplanets with orbital periods < 10 days \citep{Mayor1995}, called hot Jupiters, opened up the field of exoplanet discovery and characterization. Due to their close distance to the host star, they are among the easiest extrasolar planets to observe. However, they defy standard theories of planetary formation, which do not predict gaseous giants at such small distances from their host stars. 
 As such, hot Jupiters have inspired extensive theoretical work on exoplanet formation, thereby broadening our understanding of planetary system formation and evolution. 

There are three main proposed hot Jupiter formation theories: in situ formation \citep{Boss1997,Durisen2007,Perri1974,Pollack1996,Chabrier2014}, disk migration \citep{Goldreich1980,Lin1986,Lin1996,Ida2008,Brucalassi2014}, and high eccentricity tidal migration \citep{Rasio1996,Weidenschilling1996,Ford2006,Chatterjee2008,Wu2003,Fabrycky2007,Nagasawa2008,Katz2011,Naoz2012,Petrovich2015,Anderson2016,Storch2014a,Storch2014b,Naoz2011,Teyssandier2013,Wu2011,Hamers2017a,Xu2016,Li2014}. None of them has been proven to be the dominant channel of hot Jupiter formation based on current data.  Recent observations \citep{Quinn2012,Brucalassi2016,Quinn2014,Gaia2018,Winter2020} have revealed that the hot Jupiter rate correlates with stellar clustering. These observations provide promising evidence that stellar interactions in clusters/associations may affect the planet formation process and/or perturb orbital parameters post-formation on short timescales (i.e., shorter than the timescale for cluster disruption), and thus may play a non-trivial role in hot Jupiter formation. Neither the formation of hot Jupiters in-situ nor via disk migration are  directly correlated to the stellar clustering. On the other hand, high eccentricity migration induced by flybys is significantly affected by stellar clustering, due to its origin in dynamical interactions \citep{Malmerg2011,Wang2020b,Li2020}.

 \citet{Adams2005,Shara2016,Wang2020,Rodet2021} have shown that scatterings due to passing stars in a star cluster can change the orbits of two-planet planetary systems, producing both hot Jupiters and very distant Saturns. More generally, stellar encounters can induce external perturbations that significantly change the architecture of multiple planet systems in clusters/associations, and can thus affect the subsequent internal dynamics within the planetary system \citep{Hurley2002, Adams2006, Spurzem2009,Li2015, Malmerg2011,Cai2019, Wang2020a,Wang2020b,Li2020}. In particular, \citet{Wang2020} and \citet{ Rodet2021} have shown that close flybys in dense clusters can contribute to the formation of hot Jupiters by activating the 
  Zeipel-Lidov-Kozai
 (ZLK) effect or planet-planet scattering between two giant planets. However, the estimated hot Jupiter formation rate from this channel is lower than the observed hot Jupiter occurrence rate in clusters. 
 An additional contribution may be provided by binary star flybys or dynamics in multi-planetary systems.

In a planetary system, the angular momentum deficit is the difference between the planar 
circular angular momentum  and the total angular momentum. This quantity is conserved between 
collisions in the average system, and decreases during collisions. For systems with three or more planets, secular chaos can provide an additional mechanism to excite the eccentricity of the innermost planets to extreme values if the initial state of the system has a sufficiently large angular momentum deficit \citep{Wu2011,Lithwick2014, Hamers2017,Petrovich2019,Teyssandier2019,Oconnor2021}. This requires the planetary orbits to be mildly inclined and eccentric. However, planetary systems in the field are expected to be in nearly coplanar circular configurations due to the gas distribution during the planet formation stage. Thus, the typical angular momentum deficit of planetary systems in the field is not large enough to trigger secular chaos.

As already discussed above, stellar flybys in dense clusters can effectively change the configuration of planetary systems. In particular, 
close stellar flybys can exchange significant energy and angular momentum between a flyby star and a planetary system. If a planet survives the strong encounter without being ejected, the eccentricity and inclination of its orbit can be dramatically changed, and a significant angular momentum deficit can be acquired. In star clusters, the stellar number density is high enough for a non-negligible stellar flyby rate. If  enough multi-planet systems survive  the close encounters, there may be some multi-planet systems with sufficient angular momentum deficit to trigger secular chaos and thus activate the high eccentricity migration mechanism.

In this paper we build on our previous study \citep{Wang2020} by simulating scatterings between single stars and planetary systems hosting three planets, using our high-precision, high accuracy few-body code {\tt SpaceHub} \citep{Wang2021}.
Tidal migration and general relativistic precession are included in the long-term integration to numerically track the formation of hot Jupiters. 

The paper is organized as follows. In Sec.~2 we describe the dynamical mechanisms which result in eccentricity increases, and how we set up our numerical calculations. Sec.~3 specializes to the description of how hot Jupiters form after the onset of high eccentricity. Our numerical methods are laid out in Sec. ~4, while the results of the simulations are reported in Sec. ~5. We summarize our results and conclusions in Sec.~6.

\section{Stellar flybys on multi-planet systems}

In this section, we introduce and discuss the effects of stellar flybys on multi-planet systems, with a focus on the different mechanisms that can operate to alter the planetary orbital parameters, in particular eccentricity and inclination.

\subsection{Mechanisms of eccentricity excitation by stellar flybys}
Standard theories of planetary formation \citep{Boss1997} predict that giant planets in multi-planet
systems are typically formed in well-spaced, circular coplanar orbits. Without perturbations, most of those planetary systems will remain stable on timescales longer than several Gyr. Therefore, if hot Jupiters are not formed in-situ, their formation requires external perturbations from a companion star or a protoplanetary disk, or close planet scattering.

On the other hand, in dense stellar environments, the non-negligible rate of stellar flybys can
naturally provide strong perturbations which
significantly affect the configuration of the planetary system. It then follows that several 
dynamical processes which excite
high eccentricity can be activated. These are described in the following subsection.

\subsubsection{ZLK effect}
Stellar flybys can break the coplanar configuration of a planetary system, and thus create  inclined planet orbits. In a hierarchical triple system, the inner orbit (here, the star-Jupiter binary) can undergo ZLK oscillations  if the inclination between the inner orbit and outer orbit is in the range of [40, 140] degrees. 
During an ZLK oscillation the energy ($\propto -M_{\rm tot,in}/a$) of each orbit is conserved, but the eccentricity of the inner orbit can reach extreme values by exchanging angular momentum with the outer orbit.  The timescale for the quadrupole ZLK effect to operate is
\begin{equation}\label{eq:time-LK}
\tau_{\rm LK}=P_{\rm in}\bigg(\frac{M_{\rm tot,in}}{M_{\rm out}} \bigg)\bigg( \frac{a_{\rm out}}{a_{\rm in}}\bigg)^3(1-e_{\rm out}^2)^{3/2}\,,
\end{equation}
where $P_{\rm in}$ is the period of the Jupiter orbit, $M_{\rm tot, in}$ is the total mass of the inner orbit, and $M_{\rm out}$ is the mass of the perturber. ZLK oscillations require the triple system to be hierarchical and stable. These oscillations operate in the range in which $-\sqrt{3/5}<\cos(I)<\sqrt{3/5}$, with $I$ being the inclination between the two planets' orbits.

Previous studies \citep{Hamers2017,Wang2020,Rodet2021} have shown that stellar flybys on a system with two giant planets can efficiently activate the ZLK effect and thus increase the eccentricity of the innermost planet, eventually leading to hot Jupiter formation. In a multi-planet system, the ZLK effect no longer operates due to the existence of the other planets. However, close stellar flybys can eject the outer planets and incline the inner planet. Thus, the ZLK effect may still play a role in hot Jupiter formation in multi-planet ($\geq 3$) systems. 

\subsubsection{Secular chaos}
In a multi-planet system of three or more well-spaced planets with a sufficiently large angular momentum deficit,
secular interactions among these planets may lead to significant angular momentum exchange while not changing the orbital energy of each planet. 
The angular momentum deficit (AMD) is defined as
\begin{equation}
    AMD = \sum_{k=1}^{N}\Lambda_k(1-\sqrt{1-e_k^2}\cos i_k)\,,
    \label{eq:amd_def}
\end{equation}
where $\Lambda_k=\frac{m_kM_\circ}{m_k+M_\circ}\sqrt{G(M_\circ+m_k)a_k}$ is the circular angular momentum of the planet k, $e_k$ is the eccentricity and $a_k$ is the semi-major axis, $i_k$ is the inclination relative to the invariable plane (the plane perpendicular to the total angular momentum of the system) of the multi-planet system and $N$ is the number of the planets.
In such a system, the innermost planet can gradually increase its eccentricity and inclination towards equipartition of the conserved quantity, that is the angular momentum deficit of the whole system. This phenomenon is known as secular chaos \citep{Wu2011}. With a large enough AMD budget, the maximum eccentricity of the innermost planet can become large enough to activate the high eccentricity hot Jupiter migration mechanism.

Similarly to other secular effects, secular chaos does not change the orbital energies of the planets, but the angular momentum can be transferred among different planetary orbits. Since the AMD is conserved, the maximum eccentricity of the innermost planet that can be obtained is limited. Therefore, to activate high eccentricity migration, the AMD must be sufficiently large. If all of the AMD can be transferred to the innermost planet, in order to form a hot Jupiter with pericenter $a_J(1-e_J)<0.1$ AU, the AMD should be at least as large as
\begin{equation}\label{eq:amd}
    AMD > \Lambda_J\bigg[ 1- \bigg(\frac{a_J}{0.1AU}\bigg)^{-1/2}\cos i_J\bigg]\,.
\end{equation}
However, this condition is not enough to guarantee that the innermost orbit can obtain all the available AMD during dynamical evolution, and neither is there a timescale on which secular chaos operates. A system with large enough AMD may remain stable for an extremely long time \citep{Tamayo2020}. \citet{Teyssandier2019} found that secular chaos has some correlation with orbital secular resonances, but the parameter space that leads to secular chaos is still poorly constrained. Thus, long-time secular numerical integration is required to study when high eccentricity migration is activated via secular chaos.

\subsubsection{Other non-secular mechanisms}
Other non-secular effects like close planet-planet scatterings can also convert Keplerian shear into an angular momentum deficit, triggering high eccentricity migration \citep[e.g.][]{Rasio1996,Weidenschilling1996, Ford2006,Chatterjee2008}. In planet-planet scattering, the planets typically undergo several close encounters to grow their eccentricities to high values. However, if the orbital velocity of a planet is larger than the escape velocity at the planet's surface, the cross-section for collision is larger than the cross-section for scattering. This may lead to a planet collision rather than just eccentricity growth. \citet{Goldreich2004,Ida2013} and \citet{Petrovich2014} give the upper limit for the eccentricity from planet-planet scattering as
\begin{equation}
e_{\rm p-p}\sim\frac{\sqrt{GM_{\rm p}/R_{\rm p }}}{2\pi a_{\rm p}/P}\,,
\end{equation}
where $M_{\rm p}$ is the mass of the planet, $R_{\rm p}$ is its radius, $a_{\rm p}$ is its semi-major axis (SMA)  and $P$ its orbital period. However, from our previous study, we found that flyby-induced planet-planet scattering has a relatively low rate compared with the secular ZLK effect, because flybys alone can hardly create the proper initial conditions for  planet-planet scattering. Thus, only a very small fraction of hot Jupiters can be formed from this flyby-induced planet-planet scattering channel. 

\section{Hot Jupiter formation from high eccentricity tidal migration}

\subsection{Tidal dissipation and circularization}\label{sec:tide}
The energy of a planet can be internally dissipated by the tidal force from its host star. With tidal dissipation, a planet with an orbital period of a few days (i.e. a hot Jupiter) can be formed. The magnitude of the tidal dissipation is a sensitive function of $q=a(1-e)$. As dynamical processes like ZLK, secular chaos, and planet-planet scattering described above excite the eccentricity of the planet orbit to near unity, tidal dissipation can create hot Jupiters very efficiently.

Since all the eccentricity excitation mechanisms noted above create extremely eccentric orbits, the pericenter of the planet will be very close to its host star and hence most of the dissipation will occur at the closest approach. Therefore, for tidally-induced eccentricity migration we can adopt the dynamical model for tides. This assumes that the normal modes of the planet which arise at the closest approach will be fully dissipated before the next pericenter passage. For every  pericenter transit, the energy dissipation for a  non-rotating planet can be written as \citep{Nagasawa2008},
\begin{equation}
    \Delta E_{\rm tide} \sim \frac{1}{5\sqrt{2}}\frac{w_0\tilde{Q}^2}{\xi}{\rm exp}\bigg(\frac{-4\sqrt{2}}{3}w_0\xi\bigg)E_p\,,
\end{equation}
where $\xi = \sqrt{ \frac{mq^3}{m_\star R^3} }$, $w_0\sim 0.53 (R/R_J) + 0.68$ is the dimensionless frequency of the fundamental mode, $\tilde{Q}\sim -0.12(R/R_J)+0.68$ is the dimensionless Love number \citep{Poynting1909}, $E_p = Gm^2/R$ is the surface energy of the planet,  $m_\star$ is the mass of the star, $m$  the mass of the planet and $R$ its radius,  and $R_J$ is the Jupiter radius.

As the tide dissipates orbital energy, both the semi-major axis and the eccentricity of the highly eccentric orbit will decrease, until a circular tight orbit is formed. As mentioned above, if the normal modes of the planet excited at the closest approach can be fully dissipated before the next pericenter passage, the angular momentum of the orbit is conserved. Therefore, the final semi-major axis is $a_{\rm f} = a_0(1-e_0^2)$. The corresponding dissipation timescale for the semi-major axis  is given by
\begin{equation}\label{eq:time-tide}
     \tau_{a} \sim -\frac{a}{\dot{a}} = \frac{Gmm_\star/(2a)}{-\Delta E_{\rm tide}}P\,,
\end{equation}
where $P$ is the orbital period of the planet's orbit.

\begin{figure}
	\includegraphics[width=0.9\columnwidth]{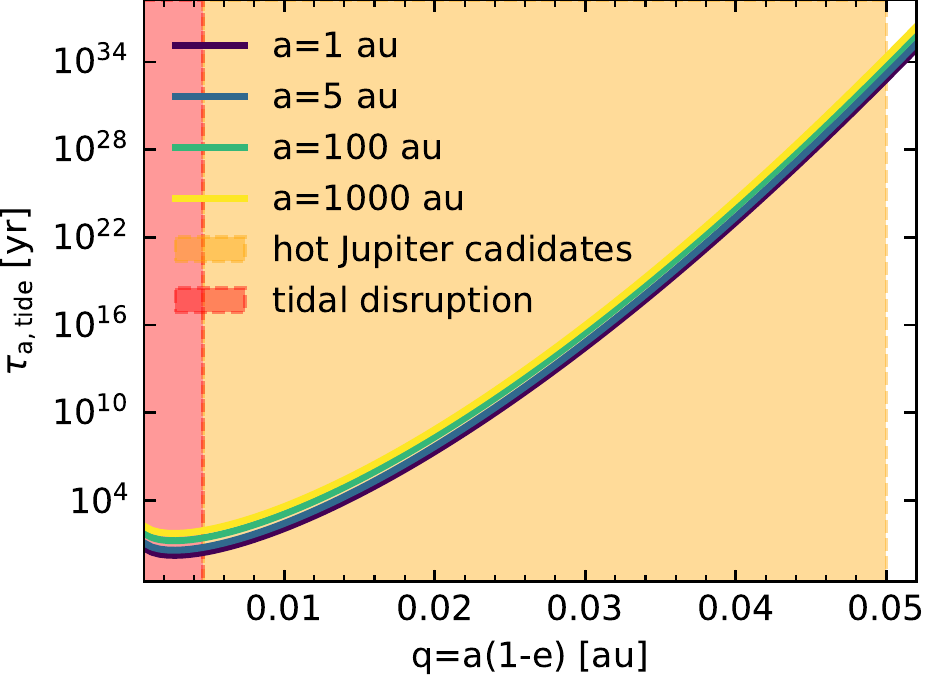}\\
    \caption{Tidal dissipation timescale as a function of $q=a(1-e^2)$ for a solar mass star and a Jupiter mass planet. The function is given by Equation~\ref{eq:time-tide}. }
    \label{fig:tidal-time}
\end{figure}

Figure~\ref{fig:tidal-time} shows the  tidal dissipation timescale for a one solar mass star and a Jupiter mass planet as a function of $q$, for different initial semi-major axes. For hot Jupiters
with periods < 10 days orbiting 1~$M_\odot$ stars,
the parameter $q$ needs to be smaller than ~0.05~au. This figure shows that, in the $q<0.05$ au region, the tidal dissipation timescale spans the very wide range 10$^3$- 10$^{34}$~years. For $q\sim0.01$~au, the tidal circularization can be completed within 10$^4$ years, while for $q\sim0.05$ au, the timescale is much longer than the age of the universe. 

Previous investigations have generally partly relied on semi-analysis methods to study hot Jupiter formation, where the high eccentricity excitation mechanism and tidal dissipation are studied independently. These methods are valid if the tidal dissipation timescale is much longer than that of the dynamical processes that drive the eccentricity to extreme values. However, if $q$ is as small as $\sim$ 0.01 au, the tidal dissipation timescale can be much shorter than that of the dynamical processes that increase the eccentricity. Since tidal dissipation decreases the eccentricity, in the small $q$ region (large $e$ region), semi-analysis methods become invalid. Tidal dissipation will suppress the eccentricity excitation for extreme $e$, and hence semi-analysis methods will overestimate the population of hot Jupiters with extremely eccentric orbits.   Therefore, tidal dissipation must be considered along with eccentricity excitation dynamics for small $q$ values, and the problem must be solved with dynamics and tidal dissipation fully coupled.

\subsection{General relativity in tidal migration}\label{sec:gr}
Short range effects like general relativistic (GR) precession can significantly suppress eccentricity excitation processes and thus cap the maximum eccentricity that can be achieved. Because the efficiency of tidal dissipation is very sensitive to the eccentricity in the high-e regime,  GR precession can tremendously change the fraction of  systems that can efficiently undergo tidal dissipation. The timescale for GR precession to operate on a binary with masses $m_1$ and $m_2$, with semi-major axis $a$ and eccentricity $e$ is
\begin{equation}
    \tau_{\rm GR} \sim \frac{a(1-e^2)c^2}{3G(m_1+m_2)}P\,,
\end{equation}
where $c$ is the speed of light and $P=2\pi\sqrt{\frac{a^3}{G(m_1+m_2)}}$ is the period of the binary. 

The gravitational wave radiation of the planet's orbit can be safely ignored during the tidal dissipation due to its long timescale, which is
\begin{eqnarray}
     \tau_{\rm GW}|_{e=0}&=&\frac{5c^5a_0^4}{256G^3m_{12}m_1m_2}\\
      &\sim=& 3.3\times10^{20}\bigg( \frac{m_1}{M_\odot}\bigg)^{-2}\bigg( \frac{m_2}{M_{J}}\bigg)^{-1} \bigg(\frac{a_{0}}{\rm au} \bigg)^4\, {\rm yr}\,,\nonumber\\
      \tau_{\rm GW}|_{e\sim 1}&=&\frac{768}{425}(1-e^2)^{7/2}\tau_{\rm GW}|_{e=0}\,,
\end{eqnarray}
where $m_1$ is the mass of the host star, $m_2$ is the mass of the Jupiter, $m_{12}=m_1+m_2$, $a_0$ is the initial semi-major axis of the Jupiter orbit and $e$ is the Jupiter orbital eccentricity.

\subsection{Tidal disruption and collision}
If the pericenter of the planet's orbit is too close to the host star, the tidal force of the star could disrupt the planet.  In this case, no hot Jupiter would form. The tidal disruption radius is 
\begin{equation}
    r_{\rm TDE} \sim \bigg(\frac{M_\star}{m_p}\bigg)^{1/3}R_p\,,
\end{equation}
where $M_\star$ is the mass of the star, $m_p$ is the mass of the planet and $R_p$ is the radius of the planet. 

We are adopting the sticky-star approximation as our criterion to decide when a collision occurs.  Hence, if the pericenter is close enough that $a(1-e) < r_{\rm coll} = R_\star+R_p$, star-planet collisions can also occur, which will destroy the planet. For a $1 M_\odot$ star with a $1 M_{\rm J}$ planet, we find $r_{\rm TDE} \sim 1 R_\odot$, which is smaller than  $r_{\rm coll}=1.1 R_\odot$. Therefore, tidal disruption of the planet is forbidden in our model and a star-planet collision will occur as the eccentricity of the planet orbit approaches unity.

\section{Methods}\label{sec:methods}
In this section, we describe our suite of numerical scattering experiments, along with the implementation of both our dissipational forces and the long-timescale integrations we perform for our post-flyby perturbed planetary systems.

\subsection{Stellar flybys on planetary systems: numerical methods}
Numerical scattering experiments of stellar flybys are performed via our high precision few-body code {\tt SpaceHub} \citep{Wang2021}. Multi-planet systems are preferred in low number density clusters like open clusters, which have relatively low velocity dispersions. Therefore, the scattering experiments are performed between flyby stars with $v_\infty = 0.366$ km s$^{-1}$ (corresponding to a velocity dispersion $\sigma=0.1$ km s$^{-1}$) and three planet systems with the innermost Jupiter-mass planet at 5 AU, an outer Jupiter mass planet at 15 AU, and a third outermost planet at three different distances, 45, 75 and 150 AU with two different masses, 1 Jupiter mass and 1 Neptune mass. For simplicity, the masses of the host stars and the flyby star are both assumed to be 1~$M_\odot$.
The phases of the three planets are generated randomly within the range [-$\pi$, $\pi$].  Hence, the orientation of the planetary system with respect to the flyby star is generated randomly within the range [-$\pi$, $\pi$], and the impact parameter at infinity of the flyby star is generated uniformly in the area of $\pi b_{\rm max}^2$,  where $b_{\rm max}$ is the maximum impact parameter, corresponding to a closest approach of $Q = 5a_3$, where $a_3$ is the semi-major axis of the third planet (located at either 45, 75 or 150 ~au).
The flyby stars are launched on hyperbolic trajectories with $r = 1500$ AU; this is a value large enough that the tidal force that the star exerts on the planetary orbit can be neglected. The trajectory from infinity to $r$ is calculated analytically. The termination time for each scattering experiment is set to two times the hyperbolic drop time,
\begin{equation}
    t_{\rm drop} = M\sqrt{\frac{-A^3}{GM_{\rm tot}}}\,,
\end{equation}
where $M$ is the corresponding mean anomaly of the hyperbolic incident orbit with $r=1500$ AU, $A$ is the semi-major axis of the hyperbolic orbit (negative) and $M_{\rm tot}$ is the total mass of the flyby star and planetary system. In total we perform $5\times 10^6$ scattering experiments for each $a_3$ and $m_3$ (mass of the third planet). This number is large enough to acquire sufficient statistics to study the various outcomes of the scattering experiment as a function of the initial conditions.

\subsection{Tidal dissipation}
After the stellar flyby, if all three planets remain bound to their host star, we perform a long-time integration on the perturbed planetary system with tidal dissipation and GR precession activated. 
In our simulations, we adopt the tidal mode from \citet{Hut1981} by implementing the tidal perturbation acceleration between stellar particles and planet particles. The tidal acceleration that the star exerts on the planet is
\begin{equation}
\mathbf{a}_{\rm tid} = -3k_{\rm AM,p}\frac{GM_*^2}{M_{\rm p}r^2}\bigg(\frac{R_{\rm p }}{r} \bigg)^5 \bigg(1+\tau_{\rm p} \frac{\dot{r}}{r}\bigg)\hat{\mathbf{r}} \,,
\end{equation}
where $r$, $R_{\rm p}$, $k_{\rm AM,p}$ and $\tau_{\rm p}$ are the particle distance, the planetary radius, the apsidal motion constant and the tidal time lag, respectively. In this work, we use $k_{\rm AM,p}=0.25$ and $\tau_{\rm p} = 0.66$~s \citep{Wang2020,Hamers2017}. Since the timescale for stellar tides to operate is much longer than that for planetary tides, we do not take stellar tides into consideration in the simulations.

\subsection{GR precession}
In our simulation, GR precession is implemented by the first order post-Newtonian pair-wise acceleration \citep{Blanchet2014}. The pair-wise acceleration can be written as
\begin{equation}
\mathbf{a}_{\rm 1PN}=\frac{GM_*}{c^2r^3}\bigg[ 4(\mathbf{r}\cdot \dot{\mathbf{r}})\dot{\mathbf{r}}  + 4GM_*\hat{\mathbf{r}} - (\dot{\mathbf{r}}\cdot \dot{\mathbf{r}})\mathbf{r} \bigg].
\end{equation}

The higher-order PN terms contribute very little to GR precession and can thus can be effectively removed from the simulations. The gravitational wave radiation of the Jupiter orbit, which corresponds to the 2.5PN correction in the N-body simulations, can be ignored as well, due to its long timescale as discussed in Section~\ref{sec:gr}.

\section{Results}\label{sec:results}
\subsection{Planetary orbits immediately post-flyby}
\begin{figure}
	\includegraphics[width=0.85\columnwidth]{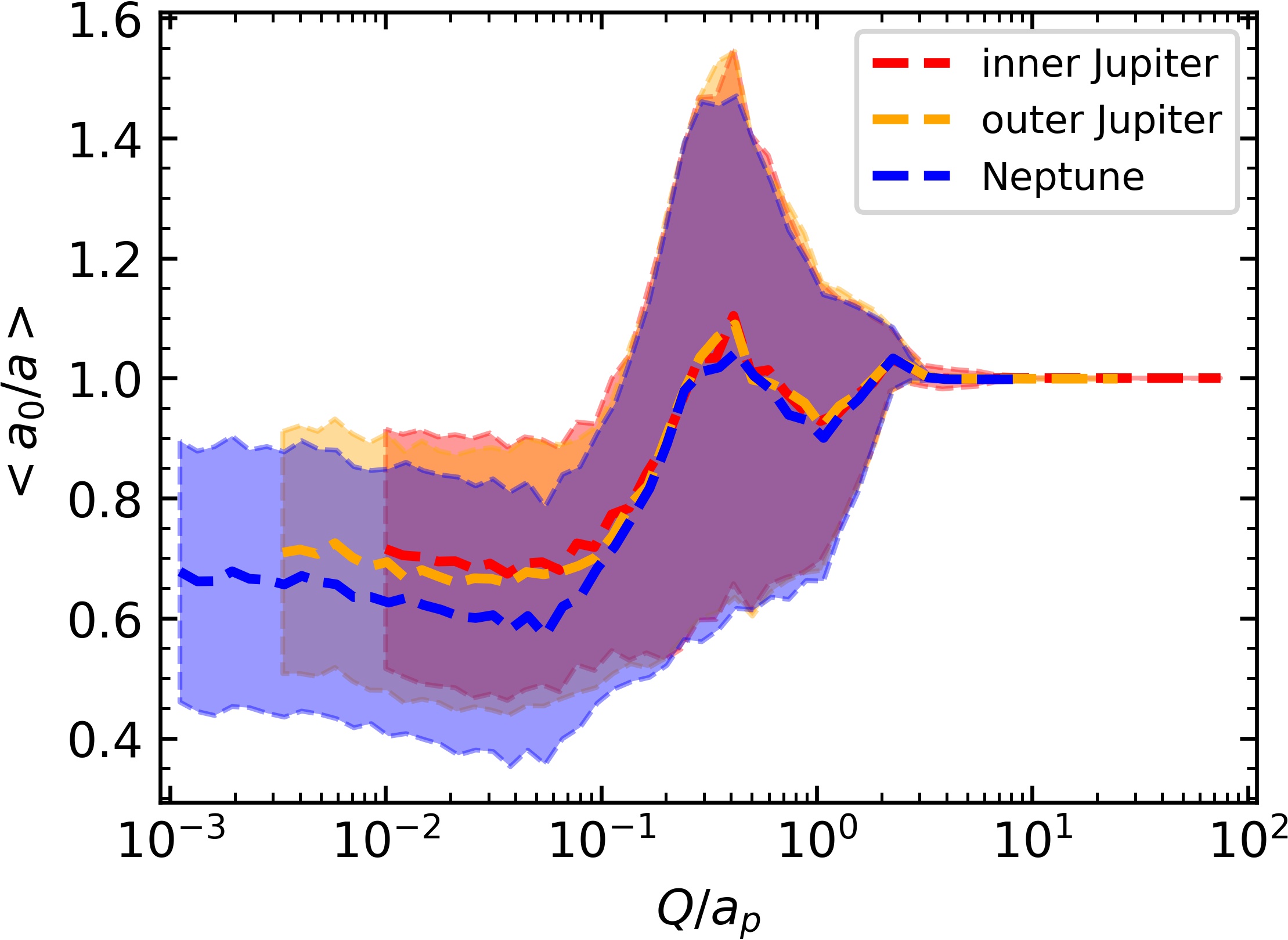}\\
	\includegraphics[width=0.85\columnwidth]{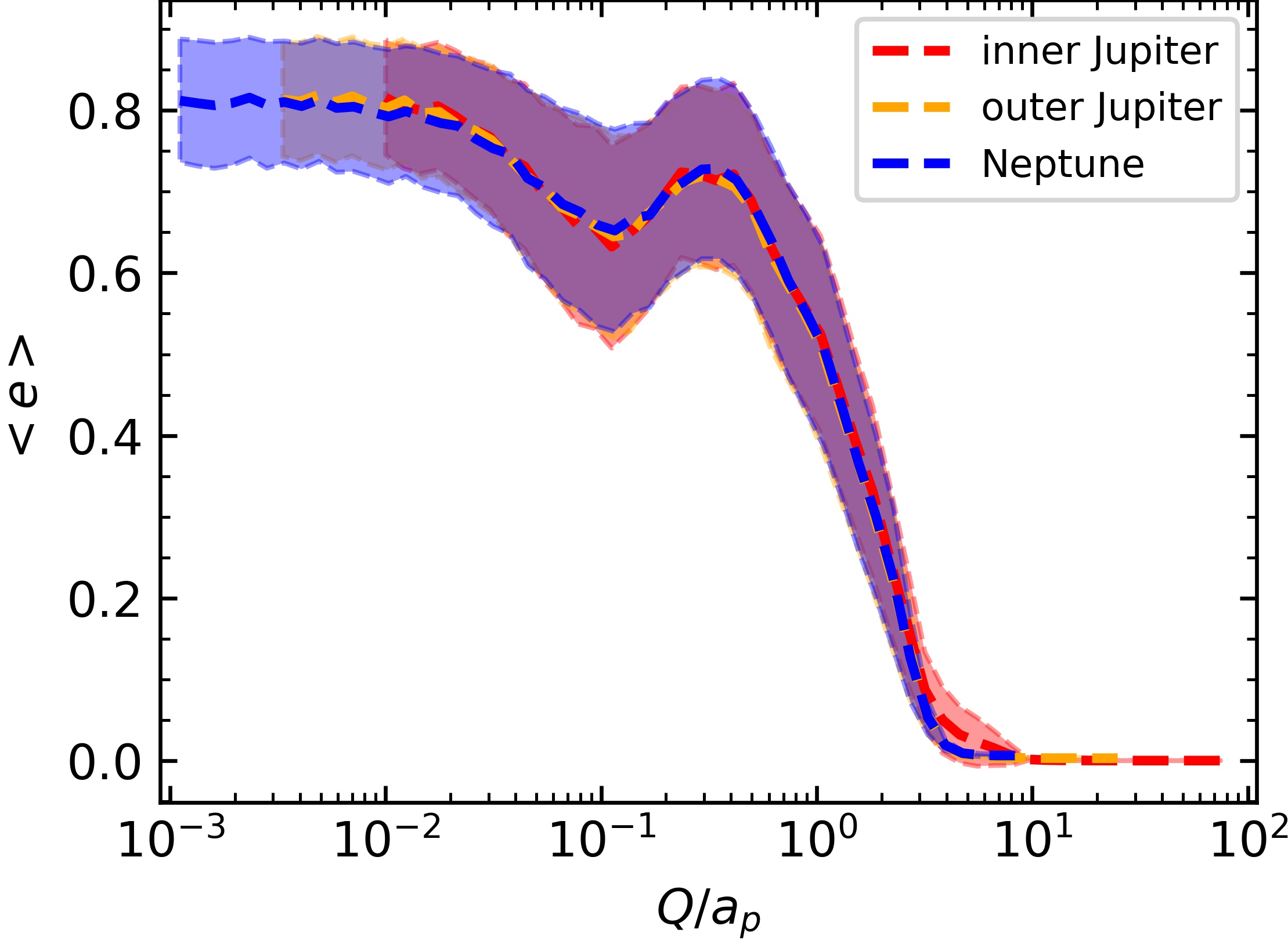}\\
	\includegraphics[width=0.85\columnwidth]{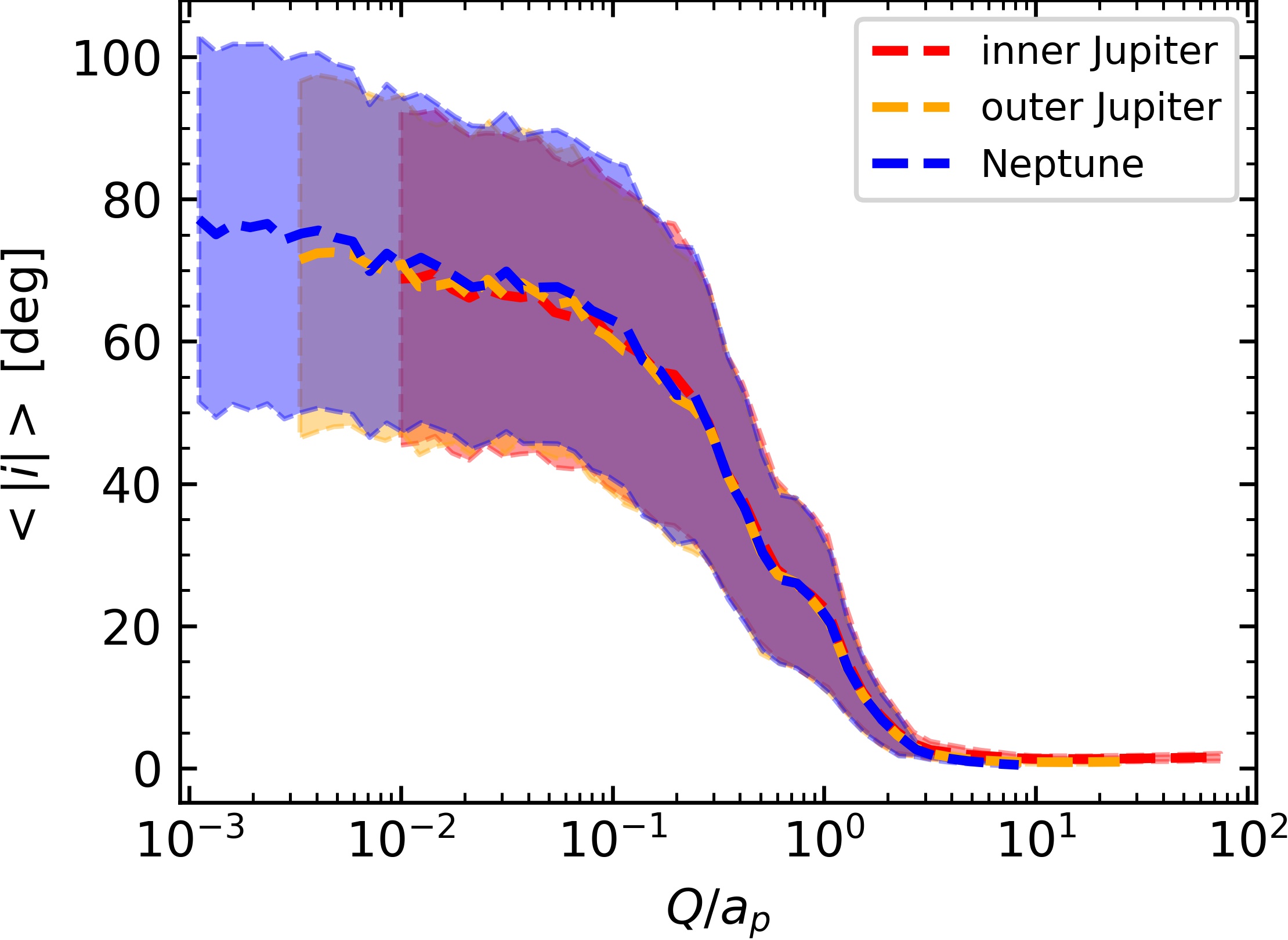}\\
    \caption{Averaged orbital properties of the planets immediately following the flyby, as a function of $Q/a_p$. The average is calculated by integrating over all the (isotropically generated) incoming angles of the perturber and all the phases of the planets.  
    The colored region shows the $1\sigma$ standard deviation. For each of the three planets,  the \textit{upper} panel shows the initial/final semi-major axis ratio, the \textit{middle} panel shows the eccentricity, and the \textit{bottom} panel shows the final inclination relative to the initial orbital plane of the planet.}
    \label{fig:post-flyby}
\end{figure}

A stellar flyby can alter the coplanar circular configuration of three-planet systems and activate secular effects like secular chaos and non-secular effects such as planet-planet scattering. The distribution of the configurations of the post-flyby planetary systems determines if certain eccentricity excitation mechanisms can be efficiently activated, and how many planetary systems can undergo high eccentricity tidal migration. 

In a three planet system, the ZLK effect can be activated if the inner two giant planets remain bound to the host star after the flyby with a sufficiently large (i.e., within the range [40, 150] degrees) flyby-induced inclination, and the outer Neptune can be ejected. If the Neptune gets ejected from the stellar flyby, this problem transitions to the one studied in \citet{Wang2020} and \citet{Rodet2021}, where only two giant planets exist before the flyby. 

Figure~\ref{fig:post-flyby} shows the orbital properties of the systems immediately after undergoing scatterings
with the flyby star as a function of $Q/a_p$. The third planet is originally placed at an orbital distance of $a_3 = $ 45 au and has a mass $m_3=0.05 M_{\rm J}$ (roughly corresponding to a Neptune mass planet) . This figure shows only the fraction of systems for which all three planets remain bound to the host star after the flyby. The upper panel shows the average ratio between the initial semi-major axis and the final semi-major axis, the middle panel shows the average eccentricity, and the bottom panel shows the average inclination. We can see that for $Q/a_p>\sim3$, the perturbation from one flyby is so small that the planetary system remains in a nearly coplanar, circular configuration. Thus, no high eccentricity excitation mechanism can be activated. However, for close flybys with $Q/a_p <\sim 2$, the eccentricity and inclination of the planetary orbits can be significantly changed, thus creating initial conditions for secular chaos, planet-planet scattering, and other eccentricity excitation mechanisms.

The results displayed in Figure~\ref{fig:post-flyby} indicate that the three planets can be treated independently during the flyby.
 This is because the average post-flyby orbital properties (dashed lines) and the standard deviation (colored region) for different planets are very similar down to $Q/a_p\sim 0.01$. 
 We will discuss this important result in more detail later on in the paper.

The scattering experiments with $a_3=$ 75 and 150~au, and with $m_3=1 M_{\rm J}$, show similar results; thus, this `independent' assumption holds true for multi-planet systems independent of the mass of the planets. This is because the mass of the planets is much smaller than the mass of the intruder star.

 For the subset of the systems for which all three planets remain bound to the host star after the flyby, we made the following categorizations.  We adopt three classes: the planet-planet scattering channel, where any two of the three planetary orbits intersect each other; the secular chaos channel, where the angular momentum deficit is larger than the criterion given by Equation~\ref{eq:amd} and the three planetary orbits do not intersect with each other; unclassified systems containing all other configurations which do not satisfy the other two criteria.
 
 \begin{figure}
	\includegraphics[width=0.85\columnwidth]{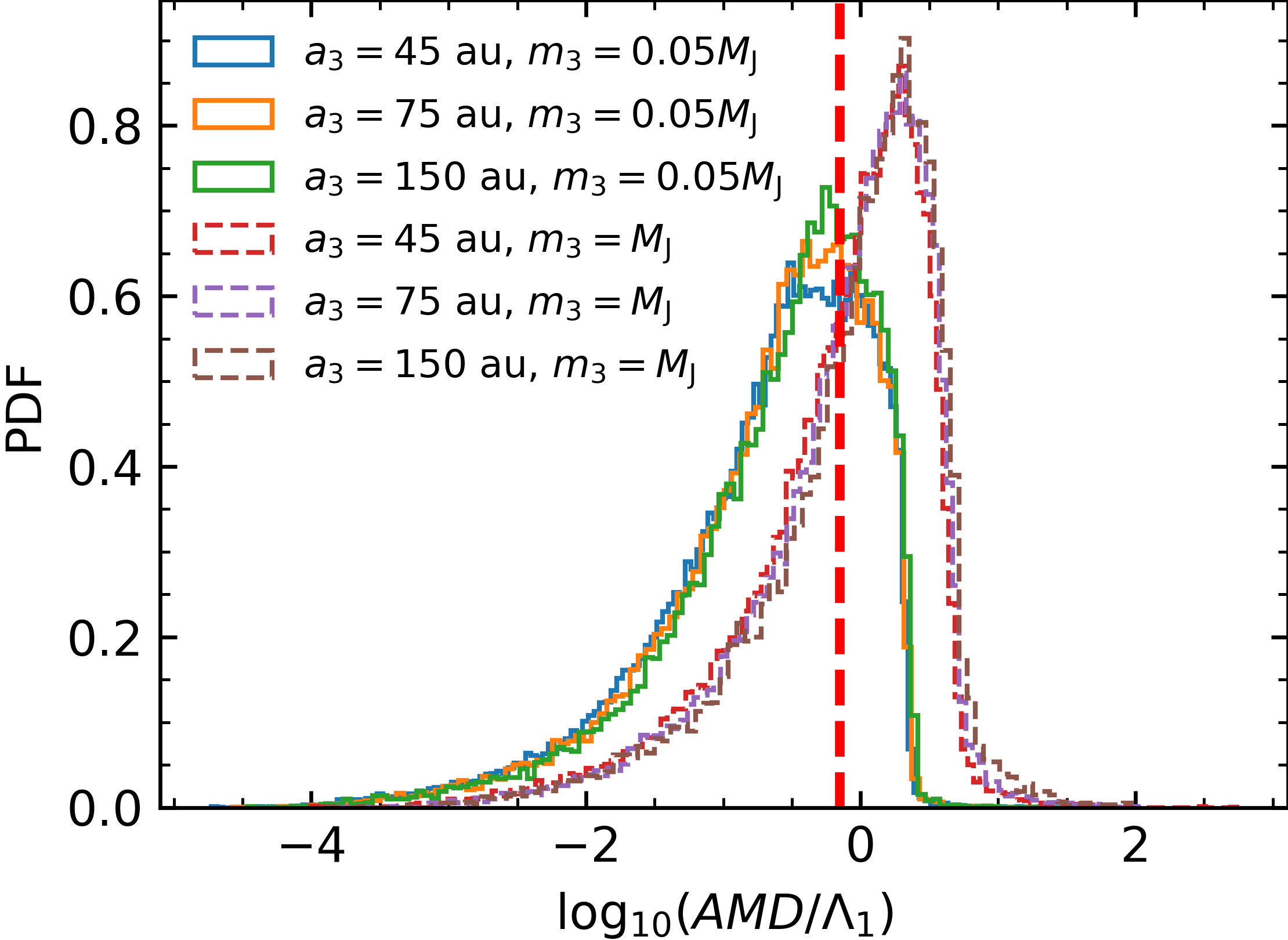}
    \caption{The distribution of the AMD (in units of $\Lambda_1$, the circular angular momentum of the innermost planet) immediately after the flyby for planetary systems with different properties (initial SMA and mass) of the outermost third planet. The two innermost, Jupiter-mass planets have initial SMA of 5 and 15~AU, respectively.
    The vertical red dashed line indicates the minimal AMD for making hot Jupiters from secular chaos. Solid lines show the distribution with a Neptune-mass outer planet, while the dashed lines show the distribution with a Jupiter-mass outer planet. The peak is around 0.5 for the Neptune-mass case,  while it is around 2 for the Jupiter mass case.}
    \label{fig:amd_tot}
\end{figure}

 Figure~\ref{fig:amd_tot} shows the distribution of the AMD of the planetary systems immediately after the flyby for different initial orbital locations and masses of the third planet: $a_3$= 45, 75 and 150 au, $m_3$= 0.05~$M_{\rm J}$ and 1~$M_{\rm J}$. 

 It can be seen that the amount of AMD imported by the scattering with the flyby star is nearly independent of the size of the planetary system but correlates with the mass of the outer planet. Systems with a more massive outer planet tend to have larger AMD after the flyby (the peak of the distribution is around 0.5 (in units of AMD/$\Lambda_1$) for the case with a Neptune-mass outer planet and around 2 for a Jupiter-mass outer planet).
  Therefore, there are more systems in the secular chaos regime after the flyby with a more massive outer planet.
 Our simulations further show that, for a Neptune-mass third planet with $a_3$= 45, 75 and 150 au, after the flyby, about 10\%, 16\% and 20\%, respectively, of the systems 
 without planet ejections will have sufficient AMD to be in the secular chaos regime, while this is the case for only 0.3\%, 0.16\% and 0.09\% of the systems undergoing planet-planet scattering. The remaining systems will be in non-overlapping, low AMD orbits. For systems with a Jupiter-mass outer planet, the fractions of systems without planet ejections which have sufficient AMD to be in the secular chaos regime are 
 18\%, 32\% and 52\% corresponding to
 $a_3$=45, 75 and 150 au, respectively, while fractions of 0.19\%, 0.11\% and 0.06\%  will undergo planet-planet scattering for those initial orbital parameters, respectively.

\subsection{Long time integration with tidal dissipation}
The tidal force is very sensitive to the pericenter of the orbit. As the secular effects such as secular chaos do not change the semi-major axis of the orbit, the tidal force becomes sensitive to the eccentricity of the orbit for a given initial semi-major axis in this scenario. As the eccentricity of the innermost planet is driven closer to unity by secular chaos, tidal dissipation becomes more efficient, and as such it acts to suppress the eccentricity growth. Because the tidal dissipation process conserves the angular momentum of the orbit, which is proportional to $a(1-e^2)$, then the final circularized semi-major axis can be fully determined by the initial angular momentum before the dissipation. However, secular chaos or planet-planet scattering will exchange angular momentum between orbits, and the timescale of the tidal dissipation can be either longer or shorter than the typical timescale for significant angular momentum exchange induced by secular chaos or planet-planet scattering. Therefore, in order to properly determine the long-term evolution of the orbital parameters of the innermost planet,
it is necessary to evolve the post-flyby systems with the inclusion of tidal dissipation throughout the full dynamical evolution, and not just at the end as is commonly done in the literature for numerical efficiency.

\begin{figure*}
	\includegraphics[width=\textwidth]{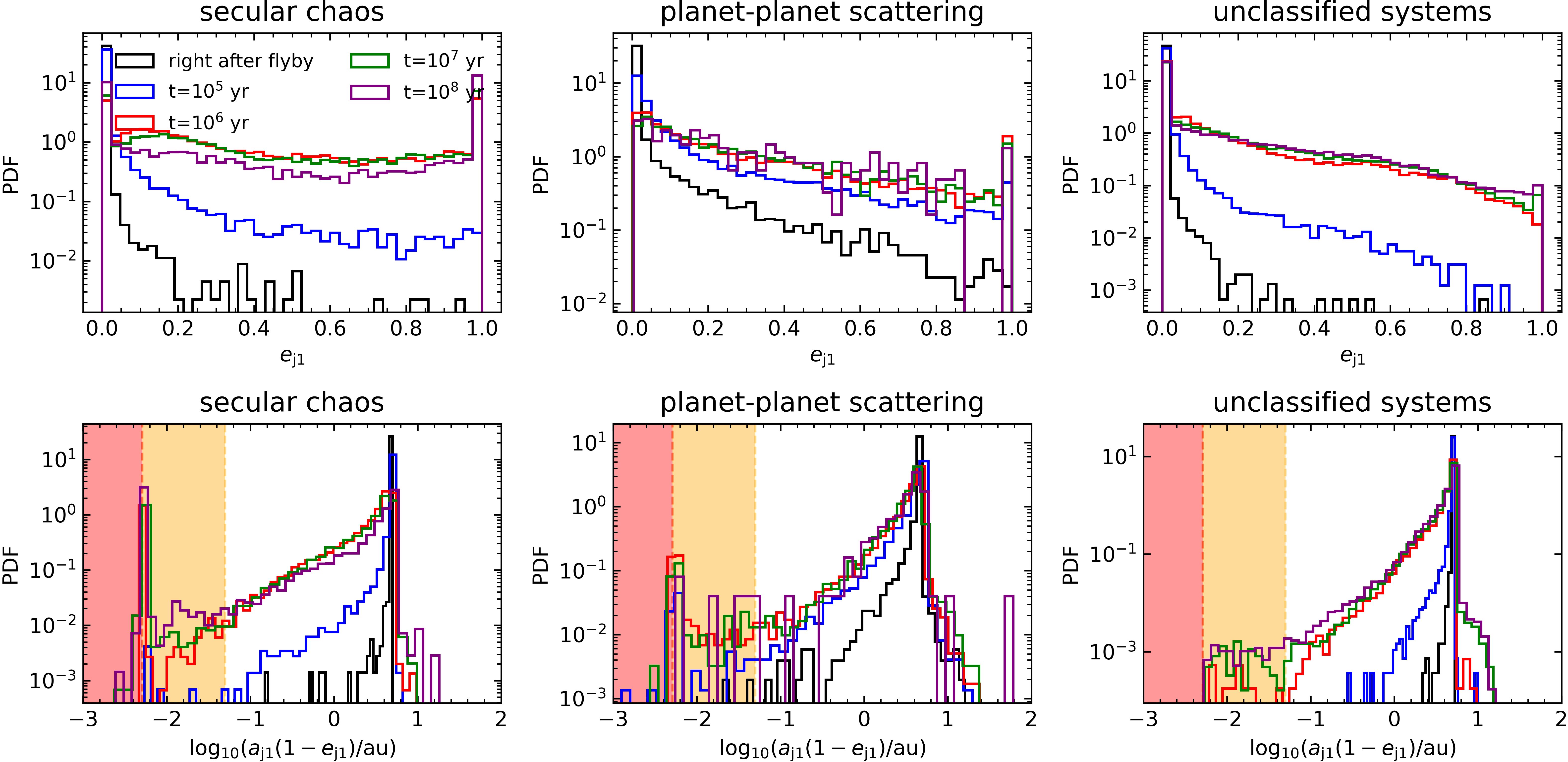}\\
    \caption{\textit{Upper} panels: Eccentricity distributions of the innermost Jupiter at different time snapshots after the flyby for systems with different post-flyby dynamical channels operating.  \textit{Bottom} panels: Pericenter distributions of the innermost Jupiter post-flyby. The systems are integrated with GR precession and tidal dissipation.  The orange regions indicate the parameter space of hot Jupiter candidates, while the salmon regions show the parameter space of star-planet collision. The initial SMAs of the three planets are 5, 15 and 45~au; the two innermost ones have a Jupiter-like mass, while the outer planet has a Neptune-like mass.}
    \label{fig:t-series}
\end{figure*}
Figure~\ref{fig:t-series} shows one of our long-term post-flyby simulations,  inclusive of tidal dissipation and GR precession. In particular, it displays the case of a planetary system with $a_3$ = 45~au and $m_3$=0.05~$M_{\rm J}$. The upper panel shows the evolution of the eccentricity for the innermost Jupiter due to different dynamical channels operating. From the distributions we can see that, right after the flyby, only a few systems obtain a non-negligible eccentricity. Most of the highly eccentric systems are found in the planet-planet scattering channel, since this requires the planets' orbits to intersect each other.  The unclassified systems and the systems in the secular chaos regime have fewer innermost planets which are 
highly eccentric post-flyby. However, the eccentricity distributions at later times
indicate that, even if secular chaos does not lead to many highly eccentric innermost planets immediately after the flyby, they are created as dynamical interactions build up over time. 

In the simulation set ups, the typical timescale for secular chaos to build up a high-eccentricity distribution is around 10$^6$ years. An enhancement of the eccentricity with time is further observed in the case of planet-planet scattering, on a typical timescale $\sim 10^5$ years. As the systems evolve in time, we can see that the statistics of planet-planet scattering become poorer, which indicates that more and more innermost Jupiters get ejected as a result of planet-planet scatterings. The bottom panels show the distributions for different times after the flyby, but for the pericenter of the innermost Jupiter. The results displayed in these panels indicate that, even though capped by GR precession, {flyby-induced} secular chaos and  planet-planet scattering are strong enough to drive the orbit of the innermost Jupiter to extremely eccentric values. Some planets may even collide with their host star. If the tidal disruption radius is outside the radius of the host star, a planetary tidal disruption event may occur.

For planetary systems with larger outer orbits ($a_3$ = 75 and 150 au) and a more massive third planet, the distribution evolves in a rather similar fashion with the timescale for secular chaos to operate just slightly shifted. The size of the planetary system is important as it will affect the total cross section of the close flyby and the probability of retaining multi-planet systems in star clusters, thus consequently affecting the hot Jupiter formation rate. However, it does not fundamentally change the way in which secular chaos makes hot Jupiters because the AMD imported by stellar flybys is similar for different sizes of planetary systems,  as shown in Figure~\ref{fig:amd_tot}.

\begin{figure*}
	\includegraphics[width=\textwidth]{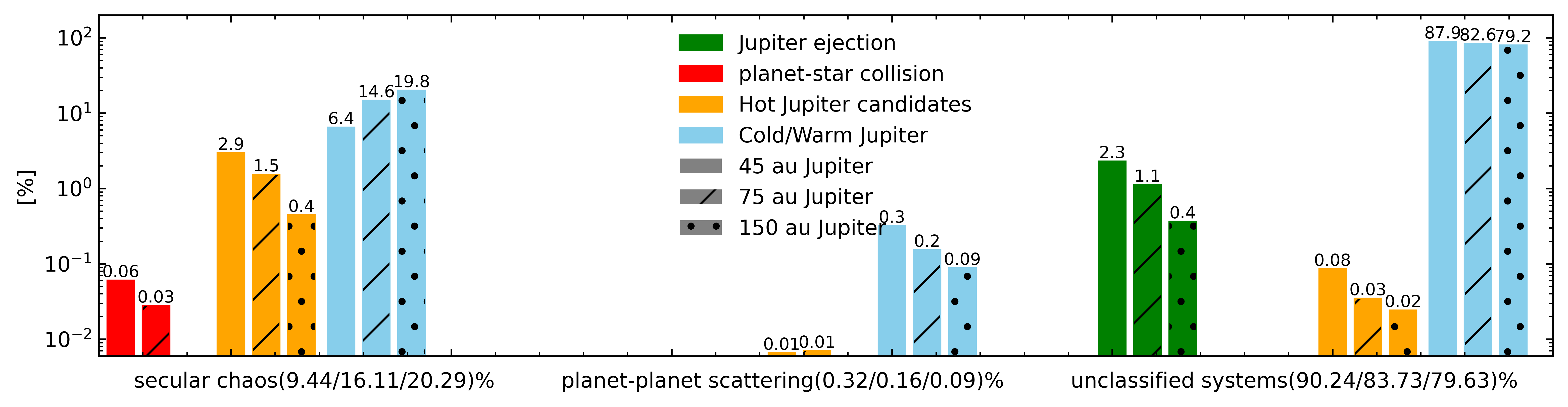}\\
	\includegraphics[width=\textwidth]{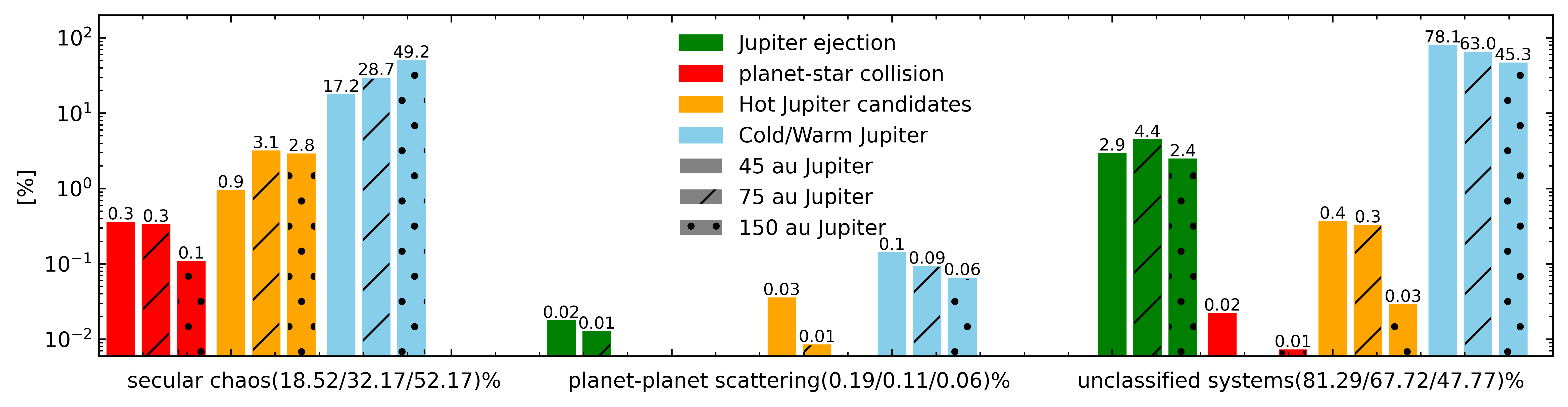}\\
    \caption{Fate of the innermost Jupiter after long-term integration (10$^8$ years) following a stellar flyby;  GR precession and tidal dissipation are included in the evolution. The simulation setup is described in Section~\ref{sec:methods}.  The upper panel shows the results for planetary systems with a Neptune-mass third planet, while the bottom panel shows the results for planetary systems with a Jupiter-mass third planet. The two innermost planets have initial SMAs of 5 and 15~AU respectively, and both have Jupiter-like mass.}
    \label{fig:frac}
\end{figure*}

Figure~\ref{fig:frac} shows the fate of the planetary systems after post-flyby long-term evolution.  Among planetary systems with an outermost planet of mass $m_3$= 0.05~$M_{\rm J}$ initially located at an orbital separation $a_3$=45~au, and which survive the close flyby without ejection,  around 10\% will be found in the secular chaos channel and around 0.3\% in the planet-planet scattering regime. The corresponding fractions for planetary systems with $a_3$=75~au and $m_3$= 0.05~$M_{\rm J}$ are 16\% and 0.16\%; and for systems with $a_3$=150~au and $m_3$= 0.05~$M_{\rm J}$ are 20\% and 0.09\%. As the size of the planetary system increases (i.e. the third planet located the furthest away from its host star), it becomes harder for the outermost planet to have a close encounter with the innermost two planets after the flyby; thus,  planet-planet scattering becomes increasingly less important. 
 
The dependence of flyby-induced secular chaos on the overall planetary size is more complex, since there are two competing effects. On one side, as the outermost planet has a larger semi-major axis, it has a larger total angular momentum, which makes it easier to acquire a larger AMD from the flyby.
However, as the outermost planet moves further away from the other two, the timescale for secular chaos to trigger extreme eccentricity in the innermost planet becomes increasingly longer. The latter effect dominates and, as a result, 
fewer hot Jupiters are formed as the semi-major axis of the outermost planet increases.
For an outer planet with an initial orbital separation $a_3$ = 45 au, about 2.9\% of the systems that survive the close flyby emerge with a hot Jupiter. This number drops to 1.5\% if the third planet is located at $a_3$=75~au, and further drops to 0.44\% for $a_3$ = 150 au.

\subsection{Hot Jupiter formation rate and comparison to the flyby-induced ZLK effect in a two-planet system}

The occurrence rate of flyby-induced hot Jupiters from secular chaos in a star cluster can be estimated as 
\begin{equation}\label{eq:ggrate}
    \Gamma =  N_{\rm three} N_\star v_\infty \theta_{\rm HJ}/ V_{\rm cluster}\,,
\end{equation}
where $N_{\rm three}$ is the number of three-planet systems in a given star cluster, $N_\star$ is the number of  stars in the cluster, $v_{\infty}$ is the relative velocity between the three planet system and the flyby star, $\theta_{\rm HJ}$  is the cross-section for flyby-induced hot Jupiter formation from secular chaos, and $V_{\rm cluster}$ is the volume of the star cluster. The cross-section $\theta_{\rm HJ}$ can be calculated from $\theta_{\rm HJ} = \theta_{\rm close} f_{\rm HJ}$, where $\theta_{\rm close}= \pi b_{\rm max}^2$ is the cross-section for close flybys, with $b_{\rm max}$ being the maximum impact parameter for which the flyby star can significantly alter the architecture of the planetary system (it usually corresponds to a closest approach of $\sim$ five times the planetary system size \citealt{Kiseleva1996}) and $f_{\rm HJ}$, the fraction of hot Jupiters  formed from close flyby-induced secular chaos, is given by the results of our long-term simulations, as described in the previous subsection.

The number $N_{\rm three}$ varies from cluster to cluster and changes as the cluster ages. If every star in the cluster is associated with three planets from the epoch of planet formation, and the interaction rate between stars is low enough that no planet gets ejected from flybys, the number of three planet systems in the cluster is equal to the number of stars in the cluster. However, if not every star hosts planets, or planets get ejected due to interactions with surrounding stars, then $N_{\rm three}$ is only a fraction of $N_\star$. Then we can assign a probability to the three-planet systems in a star cluster as
\begin{equation}
    P_{\rm three} = \frac{N_{\rm three}}{N_\star}\,.
\end{equation}
This quantity is not well constrained, neither theoretically nor observationally. We will discuss it further in the next subsection. In the following we calculate the rate of flyby-induced hot Jupiters from secular chaos {\em per three-planet system}, which is independent of the uncertain probability discussed above. This is given by
\begin{equation}\label{eq:rate}
    \gamma = \Gamma/N_{\rm three} = N_\star v_\infty \theta_{\rm HJ} /V_{\rm cluster} = n_\star v_\infty \theta_{\rm HJ}\,,
\end{equation}
where $n_\star$ is the number density of stars in the cluster.

As a reference, we can also compare the hot Jupiter rates computed here with the equivalent ones in a two giant planet system, in which case high eccentricities are excited via the ZLK mechanism following stellar flybys. In this situation,  
$f_{\rm HJ}$ was found to be about 10\% in our previous study \citep{Wang2020} of a planetary system with $a_1 = 5$~AU and $a_2 = 20$~AU in a cluster with $\sigma = 0.1$ km/s and $n= 100$~pc$^{-3}$. 

\begin{figure}
    \includegraphics[width=\columnwidth]{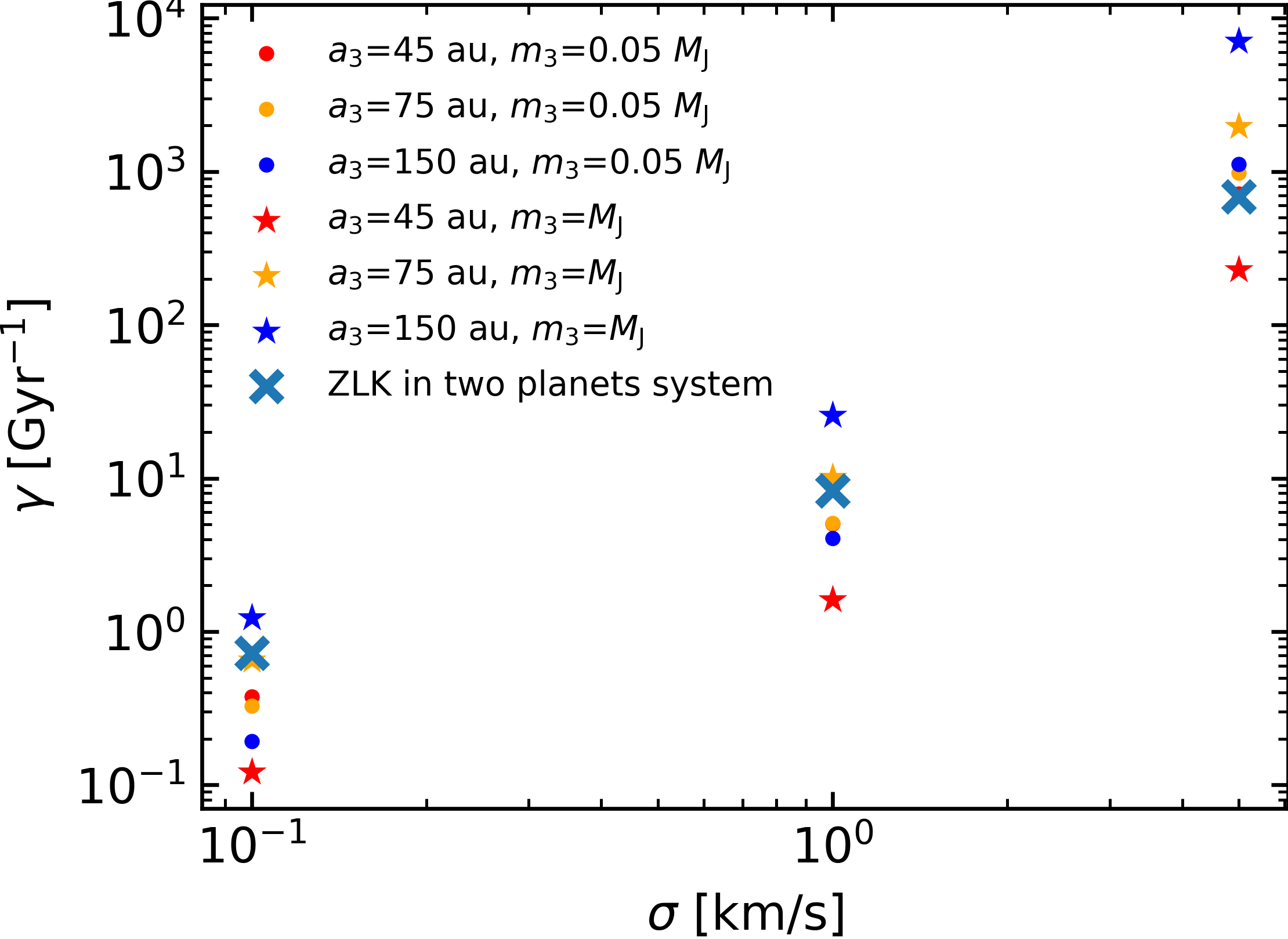}\\
    \caption{Hot Jupiter formation rate per system from flyby-induced secular chaos in star clusters. The formation rate is calculated from Equation~\ref{eq:rate}, with n-$\sigma$ pairs: [$\sigma = 0.1$ km/s, $n= 100$ pc$^{-3}$], [$\sigma = 1$ km/s, $n= 10^4$ pc$^{-3}$] and [$\sigma = 5$ km/s, $n= 10^6$ pc$^{-3}$]. The aqua-colored cross shows the formation rate from flyby-induced ZLK in two-planet systems.}
    \label{fig:rate}
\end{figure}

Figure~\ref{fig:rate} shows the hot Jupiter formation rate from flyby-induced secular chaos per three-planet system and from flyby-induced ZLK per two-planet system.  Our long-term simulations indicate that, for systems with a Neptune-mass third planet, as $a_3$ increases, $f_{\rm HJ}$ decreases, due to the inefficiency of secular chaos with a far away third planet. However, the cross-section $\theta_{\rm close}$ of a close flyby increases as $a_3$ becomes larger. The increase in $\theta_{\rm close}$ in low velocity dispersion clusters is not very significant because, with low velocity dispersion and hence small $v_\infty$, the cross section of close flybys is determined more by gravitational focusing, while with high velocity dispersion, the cross-section of close flybys is determined more by the geometric cross section of the planetary system. As a result, in the high velocity dispersion regime, a planetary system with a larger size yields a higher rate of hot Jupiter formation from flyby-induced secular chaos due to its larger geometric cross section. On the other hand, in the low velocity dispersion regime in which the geometric cross-section of the planetary system does not much affect the close-flyby cross section, larger planetary systems, which have lower $f_{\rm HJ}$, also yield a lower $\gamma$.

 For systems with a Jupiter-mass third planet, the fraction $f_{\rm HJ}$ does not vary significantly with the planetary size $a_3$. Thus, $\gamma$ is mostly determined by $\theta_{\rm close}$. Therefore, larger sized planetary systems, which are characterized by a larger $\theta_{\rm close}$, give a larger rate $\gamma$, nearly independent of the value of the velocity dispersion.

We remark that the rate $\gamma$ of hot Jupiter formation represents the rate per system, given a  three-planet or two-planet system. To estimate the global formation rate in star clusters, $\Gamma$, we need to include information also on the cluster population of  three- and two-planet systems. In clusters with high stellar number density (corresponding to higher $\sigma$), multi-planet systems get disrupted more easily. Therefore, multi-planet systems tend to be less populated in denser clusters, giving a lower number for $N_{\rm three}$. In the next subsection, we will make an estimate for the population of multi-planet systems in star clusters and calculate the convolved hot Jupiter formation rate from both flyby-induced secular chaos and ZLK.

\subsection{Survivability of multi-planet systems in dense clusters and implications for the hot Jupiter formation rate }\label{sec:thre-planets}

The existence of a population of multi-planet systems in clusters is essential to flyby-induced high eccentricity migration, since this determines how many systems can be used as the source of flyby-induced hot Jupiter formation. \citet{Cai2019}) studied the survivability of multi-planet systems in dense clusters with sophisticated N-body cluster simulations. However, due to the expensive simulation cost, only a few cases were explored with a limited range of velocity dispersions and number densities of the cluster. 
Here, to study the population of  multi-planet systems in a cluster, we build a semi-analytical model to estimate the occurrence rate of multi-planet systems which allows us to explore a much wider range of cluster parameters.

In clusters, both long-range perturbations from remote stars and short-range flybys from close stars can change the architecture of a planetary system via energy and angular momentum exchange. However, long distance perturbations are in the secular regime in which only angular momentum is exchanged between planets and remote perturbers. There is no energy exchange in the long-distance perturbations, since significant energy exchange requires direct interactions \citep{Leigh16b,StoneLeigh19}. 
Thus, since long distance perturbations do not eject planets, they do not affect the survivability of multi-planet systems directly. Only close flybys can import large quantities of energy to ionize planets \citep{Leigh16b,StoneLeigh19}.  

Since the timescale of close flybys is much shorter than the timescale of long distance perturbations, and long distance perturbations cannot eject planets, we can thus safely ignore long distance perturbations in close flyby calculations. To calculate the probability of a multi-planet system with N planets surviving a close flyby, we build a toy model in the following way. 

For an isotropic close flyby between a single star and a multi-planet system with impact parameter $b$ and velocity at infinity $v_\infty$, we let the ejection probability of the i-th planet be $p_i(b, v_\infty, m_i, a_i, M)$, where $m_i$ and $a_i$ are the mass and semi-major axis of the target planet, respectively, and $M$ is the mass of the star.
Then, the planet has an overall probability 
\begin{equation}
    P_i(v_\infty, m_i, a_i, M) = \frac{\int_0^{b_{\rm ej,i}}2\pi b p_i(b, v_\infty, m_i, a_i, M)db}{\int_0^{b_{\rm ej,i}}2\pi bdb}
\end{equation}
of being ejected as a result of a close stellar flyby, where $b_{\rm ej,i}$ is the maximum impact parameter for which the flyby star can eject the planet i, which usually corresponds to the distance of closest approach $Q\sim 3 a_i$ \citep{Kiseleva1996}.
During the lifetime $T_{\rm cluster}$ of a cluster with number density $n$ and velocity dispersion $\sigma$, a number 
\begin{equation}
    N_{\rm i} = \pi b_{\rm ej,i}^2 n\sigma T_{\rm cluster}
\end{equation}
of {\color{Plum}ejection} flybys may occur for each individual planetary system. Thus, for an individual planet, the survival probability during the lifetime of the cluster is
\begin{equation}
 (1-P_i)^{N_{\rm i}}\,.
\end{equation}
Hence the probability of a three-planet system surviving a close flyby during the lifetime of the cluster is
\begin{equation}\label{eq:survive}
   P_{\rm three}(v_\infty, m_1, m_2, m_3, a_1, a_2, a_3, M) = \Pi_1^3(1-P_i)^{N_{\rm i}}\,.
\end{equation}
Note that, since $m_i \ll M$, the planet can be treated as a test particle during the close flyby and the ejection probability $p_i$ is effectively independent of $m_i$. Since most of the energy and angular momentum exchange occurs at the distance of closest approach, $b$ and $a_i$ can be reduced to $\tilde{Q}=Q/a_i$, where $Q$ is the distance at closest approach between the flyby star and the centre of mass of the planetary system.

Then, the ejection probability function can be written as $p_i(v_\infty, \tilde{Q}, M)$ and the total ejection fraction integrated over $\tilde{Q}$ can be expressed as 
\begin{equation}
    P_i(v_\infty, v_i, M) = \frac{\int_0^{\tilde{Q}_{\rm ej}}2\pi (\tilde{Q}+{v_i^2}/{v_\infty^2}) p_i(v_\infty, \tilde{Q}, M)d\tilde{Q}}{\int_0^{\tilde{Q}_{\rm ej}}2\pi (\tilde{Q}+{v_i^2}/{v_\infty^2})d\tilde{Q}}
\end{equation}
where $\tilde{Q}_{\rm ej}\sim3$ and $v_i=\sqrt{GM/a_i}$ is the orbital velocity of the planet. Using for the impact parameter distribution
\begin{equation}
 b^2 = a_i^2\left(\tilde{Q}^2+2\tilde{Q}\frac{GM}{a_iv_\infty^2}\right)=a_i^2\tilde{Q}^2+2\tilde{Q}\frac{GMa_i}{v_\infty^2}
\end{equation}
we can write
\begin{equation}
    N_{\rm i} \sim= \pi n\sigma T_{\rm cluster}\left(9a_i^2 + 6\frac{GMa_i}{v_\infty^2}\right)\,.
\end{equation}
In multi-planet systems, if the ejection of each planet can be treated independently during the close flyby, then $p_1=p_2=p_3=...=p_i$. 

To verify the conditions under which this assumption holds, and hence to compute $p_i(v_\infty, \tilde{Q}, M)$, we perform scattering experiments between representative  planetary systems and single flyby stars. More specifically, the planetary systems consist of three planets: two inner planets of Jupiter mass, and one outer planet of Neptune mass. The host star and the flyby star are assumed to be of 1~$M_\odot$. We explore three different initial planetary system configurations: a compact configuration in which the planets have semi-major axes $a_1 = 5$ au, $a_2 = 13.5$ au and $a_3 = 31.25$ au; a standard configuration with $a_1 = 5$ au, $a_2 = 15$ au and $a_3 = 45$ au; a wide configuration with $a_1 = 5$ au, $a_2 = 20$ au and $a_3 = 80$ au. All planetary orbits are assumed to have zero eccentricity initially.
We further perform scattering experiments with three different velocity dispersions, that is $\sigma=0.1$ km/s, $1$ km/s and $10$ km/s.

\begin{figure*}
	\includegraphics[width=\textwidth]{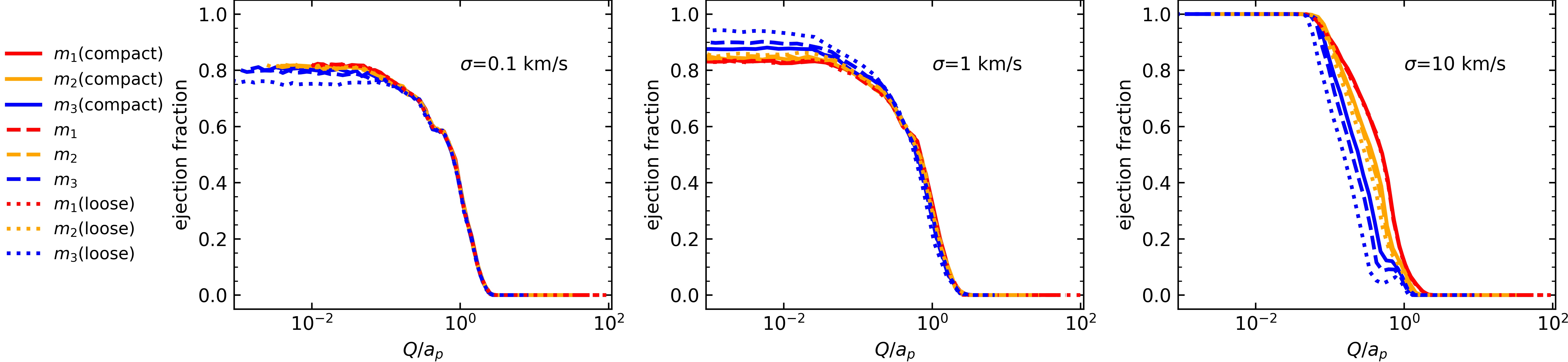}
    \caption{ Ejection fractions of a planet as a function of $Q/a_p$ in 5-body scattering experiments, where $Q$ is the distance of closest approach of the flyby star and $a_p$ is the semi-major axis of the corresponding planet. The compact  configuration has planets with semi-major axes of 5, 12.5 and 31.25~au, the standard configuration has planetary semi-major axes of 5, 15 and 45~au, while the loose configuration has the planets at distances of 5, 20, 80~au.  All planetary orbits start from zero eccentricity.
    The {\em left} panel shows the case with cluster velocity dispersion $\sigma=0.1$ km s$^{-1}$, the {\em middle} panel displays the results for $\sigma=1$ km s$^{-1}$ and the {\em right} panel shows the case $\sigma=10$ km s$^{-1}$.}
    \label{fig:ejection-fraction}
\end{figure*}

Figure~\ref{fig:ejection-fraction} shows the ejection probability function $p_i(v_\infty, \tilde{Q}, M = 1M_\odot)$ as a function of $\tilde{Q}$ for each individual planet in the three planet system, with each panel displaying the results for one particular value of the velocity dispersion. 
The figure indicates that, for different planet configurations, that is compact, standard and wide, the ejection probability functions are almost identical down to $\tilde{Q} \sim 0.1$ for velocity dispersions $\sigma = 0.1$ km/s and $\sigma = 1$ km/s. Thus in those regions the ejections of planets can be treated individually and safely scale with planet semi-major axis $a_i$. The survivability of a multi-planet system in a cluster can then be readily estimated via Equation~\ref{eq:survive}.

\begin{figure}
	\includegraphics[width=0.9\columnwidth]{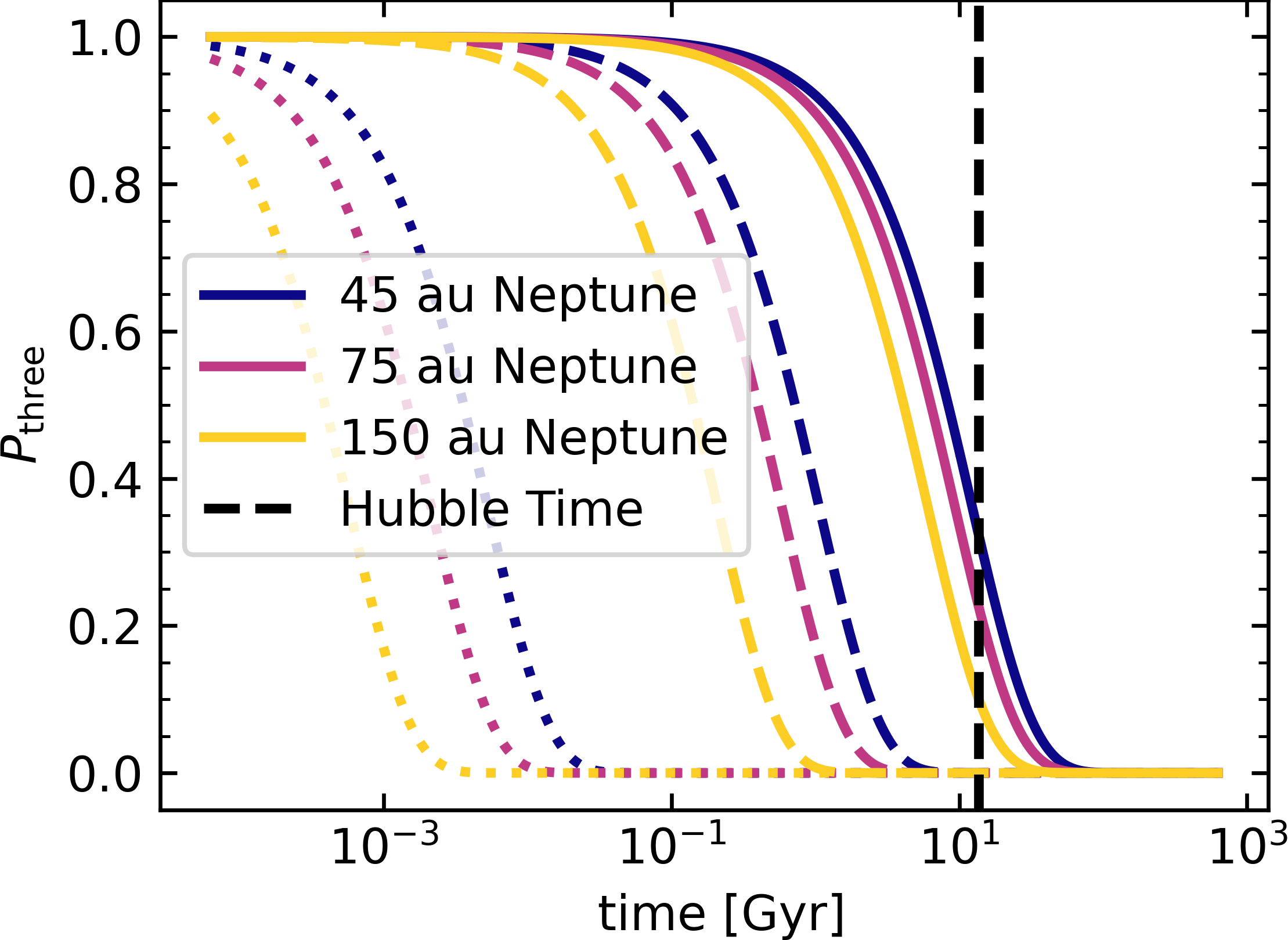}
    \caption{The probabilities of finding a three-planet system in a star cluster as a function of time. The compact configuration refers to $a_1 = 5$ au, $a_2 = 15$ au and $a_3 = 45$~au, the standard configuration refers to $a_1 = 5$ au, $a_2 = 15$ au and $a_3 = 75$~au,  while the wide configuration is the one with $a_1 = 5$ au, $a_2 = 15$ au and $a_3 = 150$~au. The solid lines show the probability in an open cluster with number density $n = 100$ pc$^{-3}$ and velocity dispersion $\sigma=0.1$~km~s$^{-1}$. The dashed lines show the probability in young massive clusters with number density $n = 10^4$ pc$^{-3}$ and velocity dispersion $\sigma=1$ km~s$^{-1}$. The dotted lines show the probability in globular clusters with number density $n = 10^5$ pc$^{-3}$ and velocity dispersion $\sigma=5$ km~s$^{-1}$. }
    \label{fig:survive3}
\end{figure}

Figure~\ref{fig:survive3} shows the probability  $P_{\rm three}$ as a function of time for different values of $n$ and $\sigma$, and for our three representative planetary configurations.  It should be noted that the number density $n$  and the velocity dispersion $\sigma$ are not generally independent. However their correlation varies as the star cluster evolves from non-virialized to viralized. Ideally, an exact calculation would require a choice of the ($n$,$\sigma$) pair for the probability calculation for each given age of the star cluster. However, this is very cluster-dependent, and hence beyond the purpose of our work, as we aim at deriving more general results.

As the figure shows, in globular clusters even the most compact planetary systems cannot survive completely unaffected after 1~Gyr. Some of their planets (most likely the outer one) will be ejected by the frequent close flybys. In open clusters, after $\sim 10$~Gyr of stellar interactions, a three-planet systems will have a probability of $\sim 40$\%  to survive. Since Galactic open clusters tend to dissolve in a much shorter time than 10 Gyr, a three planet system will have an even higher chance of survival. On the other hand, globular clusters are usually very old, which makes it very difficult for a three planet system to survive. 

With this multi-planet system population calculation in star clusters and $\Gamma$ derived via Equation~\ref{eq:rate}, we can estimate the total flyby-induced hot Jupiter formation rate of a given cluster. If the probability of finding a three planet system for a given star in a cluster after the flyby is $P_{\rm three}$ and the total number of stars is N$_\star$, then, the flyby-induced secular chaos hot Jupiter formation rate of the cluster {given by Equation~\ref{eq:ggrate} can be rewritten as,}
\begin{equation}\label{eq:grate}
    \Gamma = P_{\rm three} N_{\star} \gamma
\end{equation}
where we assume that the binary fraction in the host star cluster is relatively low(<10\%, \citep[e.g.][]{Hurley2007,Leigh11}). The total rate for flyby-induced ZLK can be calculated in a similar way.

\begin{figure*}
	\includegraphics[width=0.63\columnwidth]{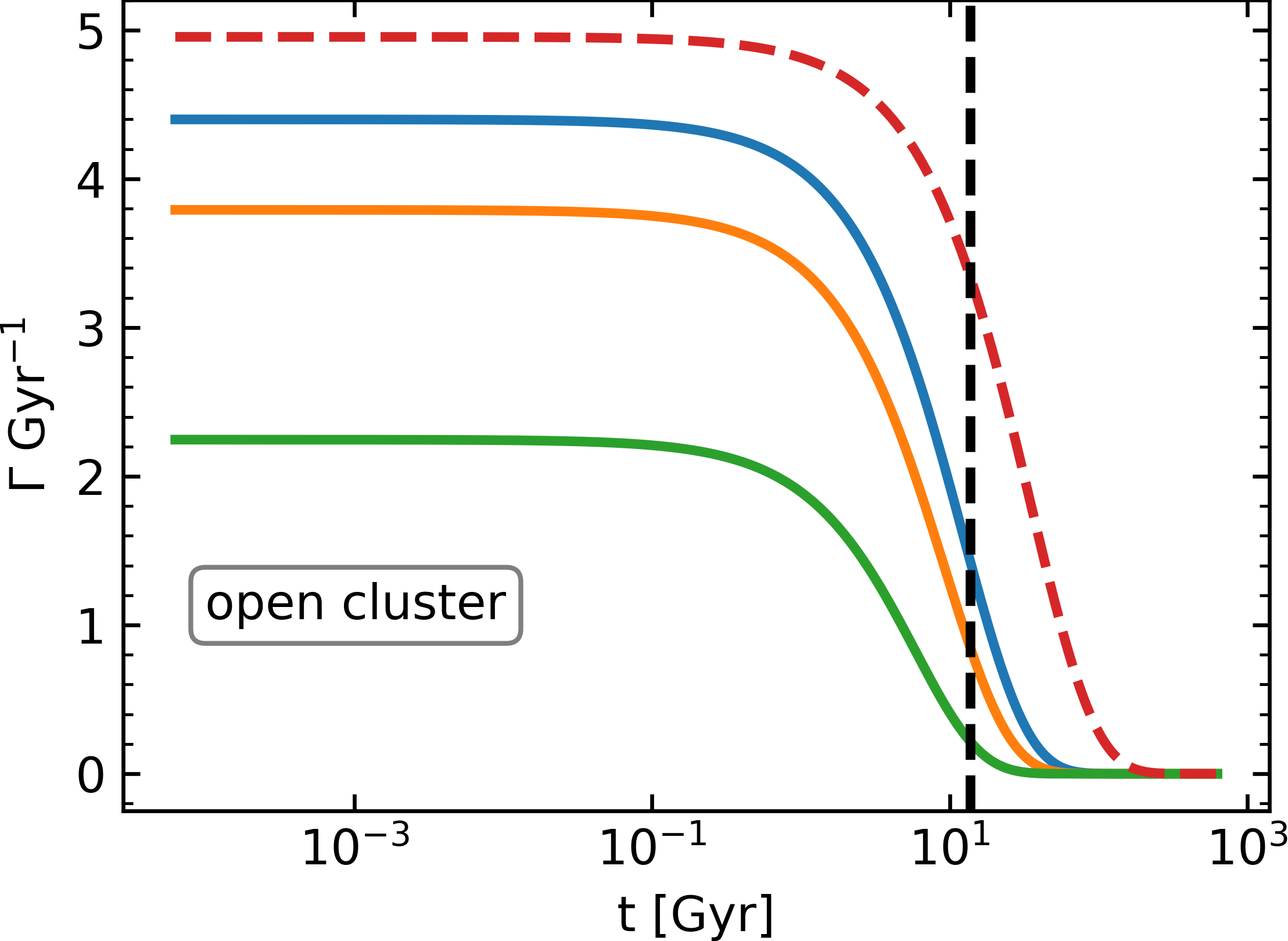}
	\includegraphics[width=0.66\columnwidth]{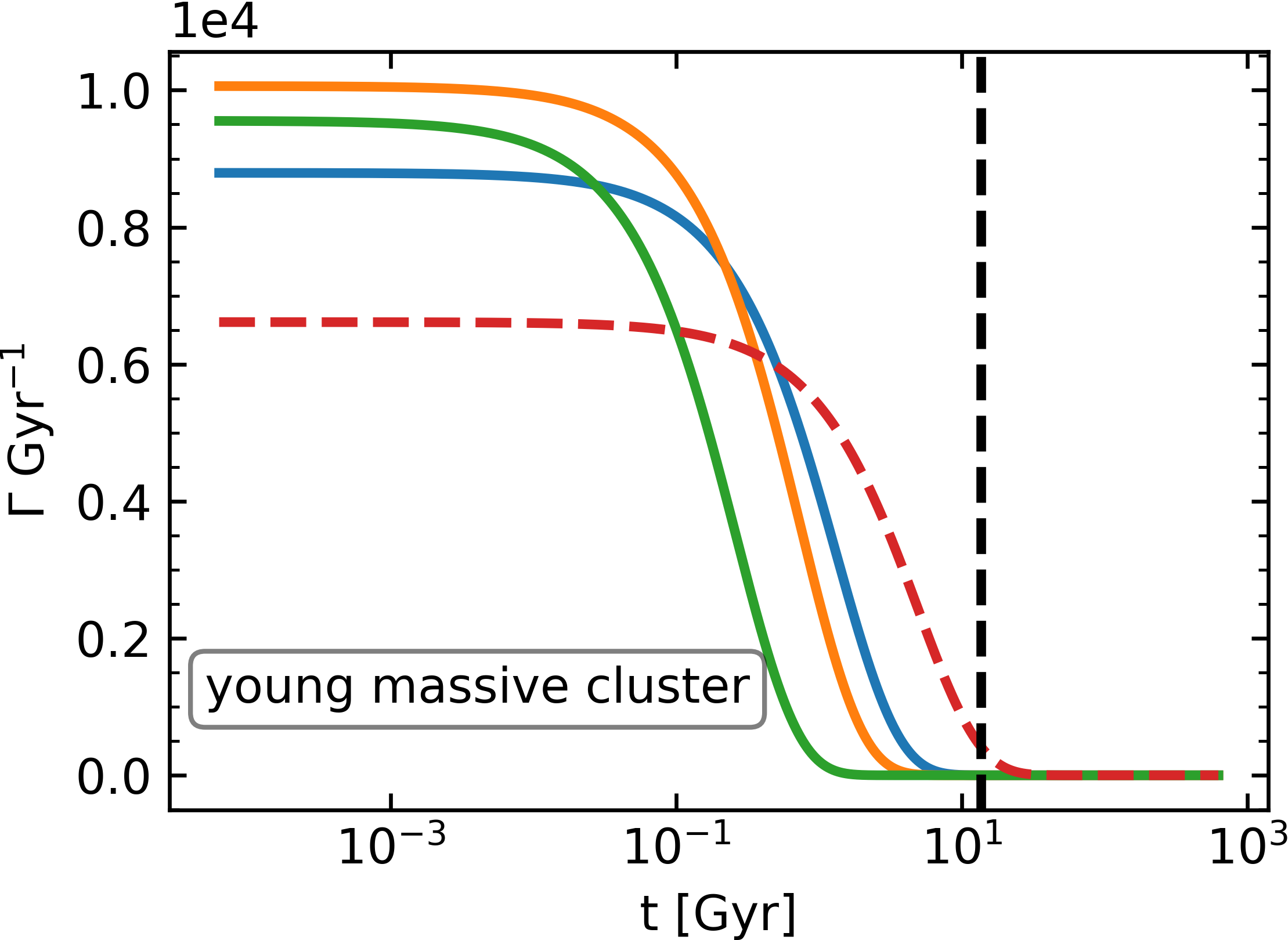}
	\includegraphics[width=0.66\columnwidth]{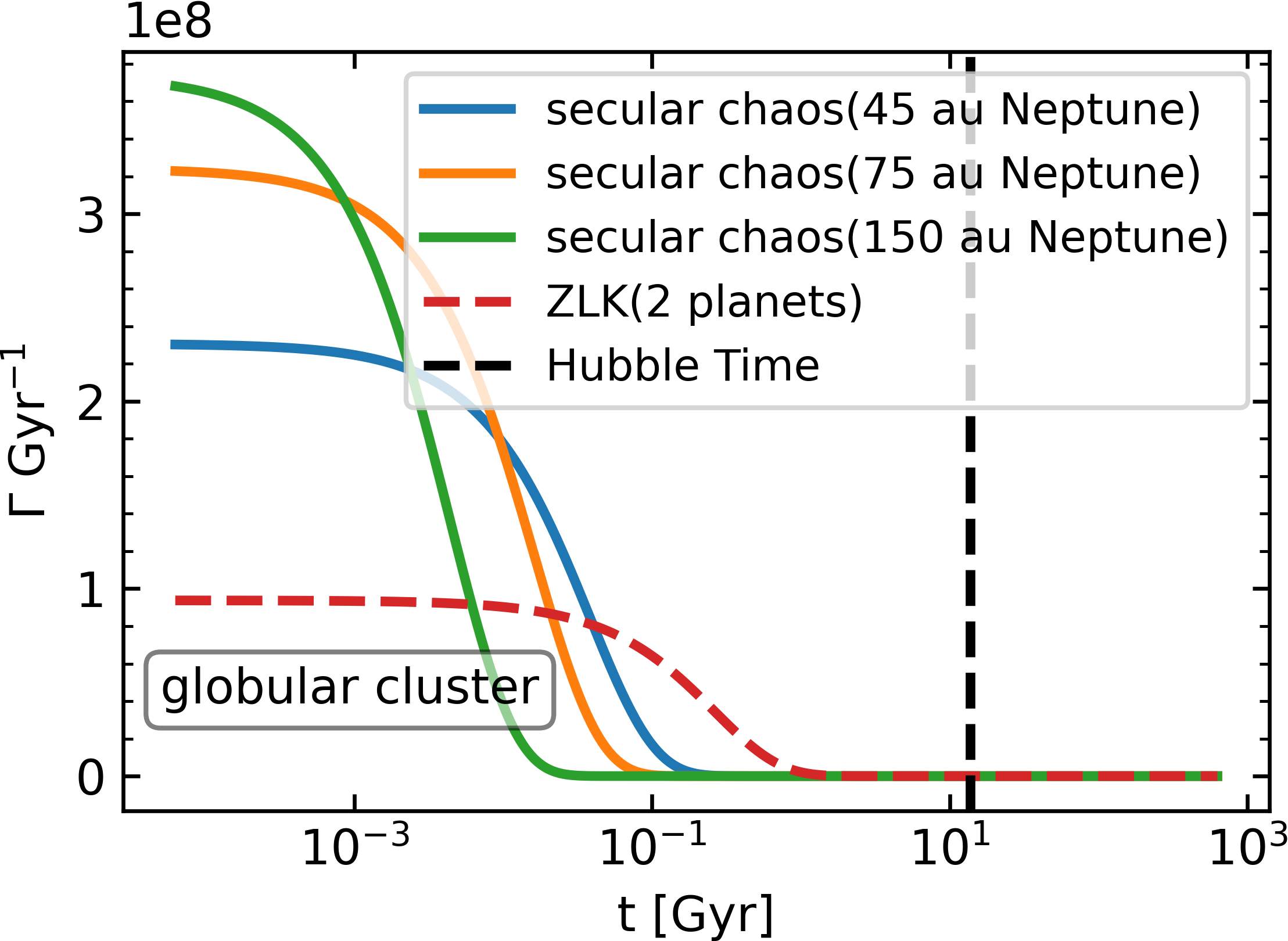}\\
	\includegraphics[width=0.66\columnwidth]{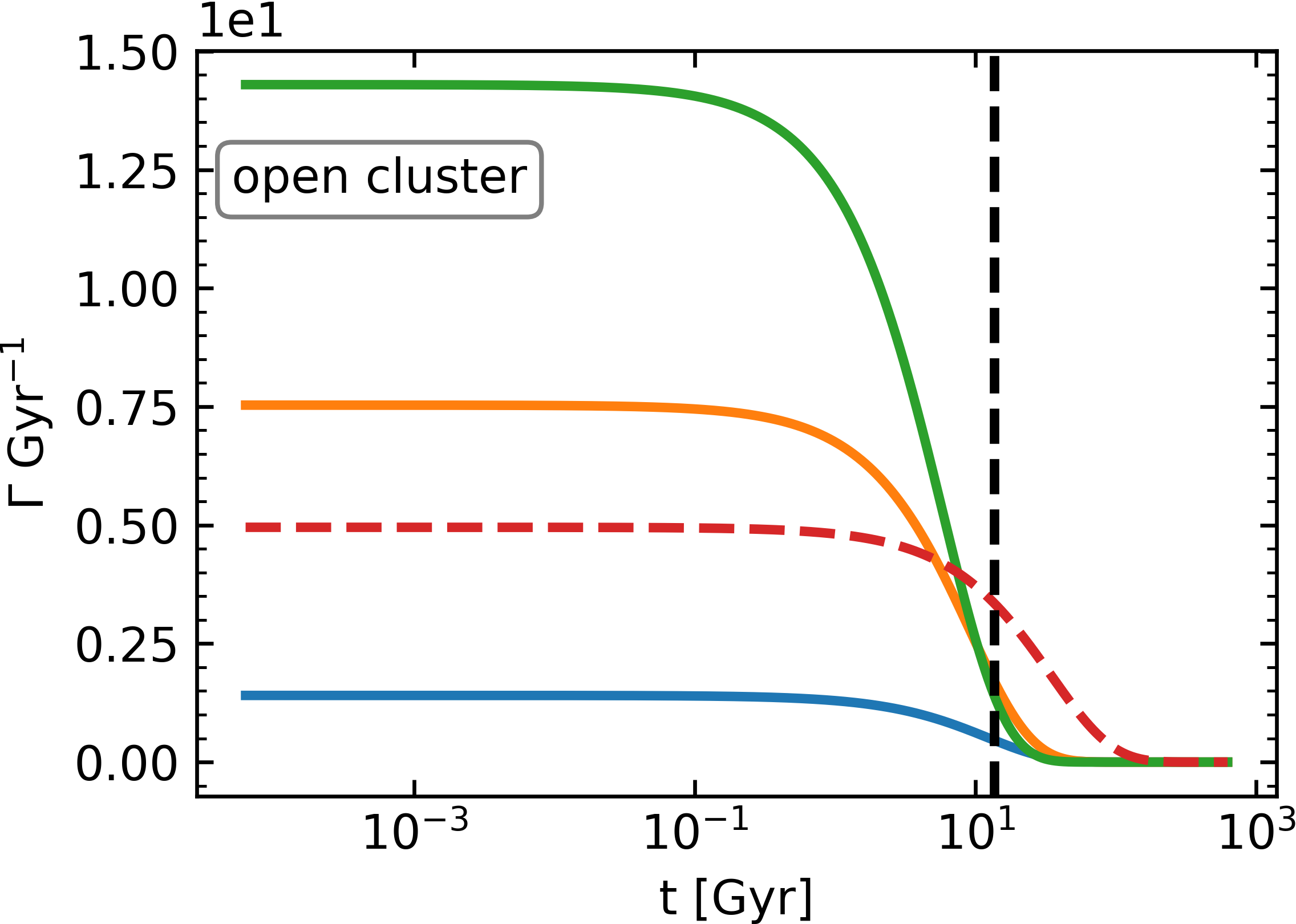}
	\includegraphics[width=0.66\columnwidth]{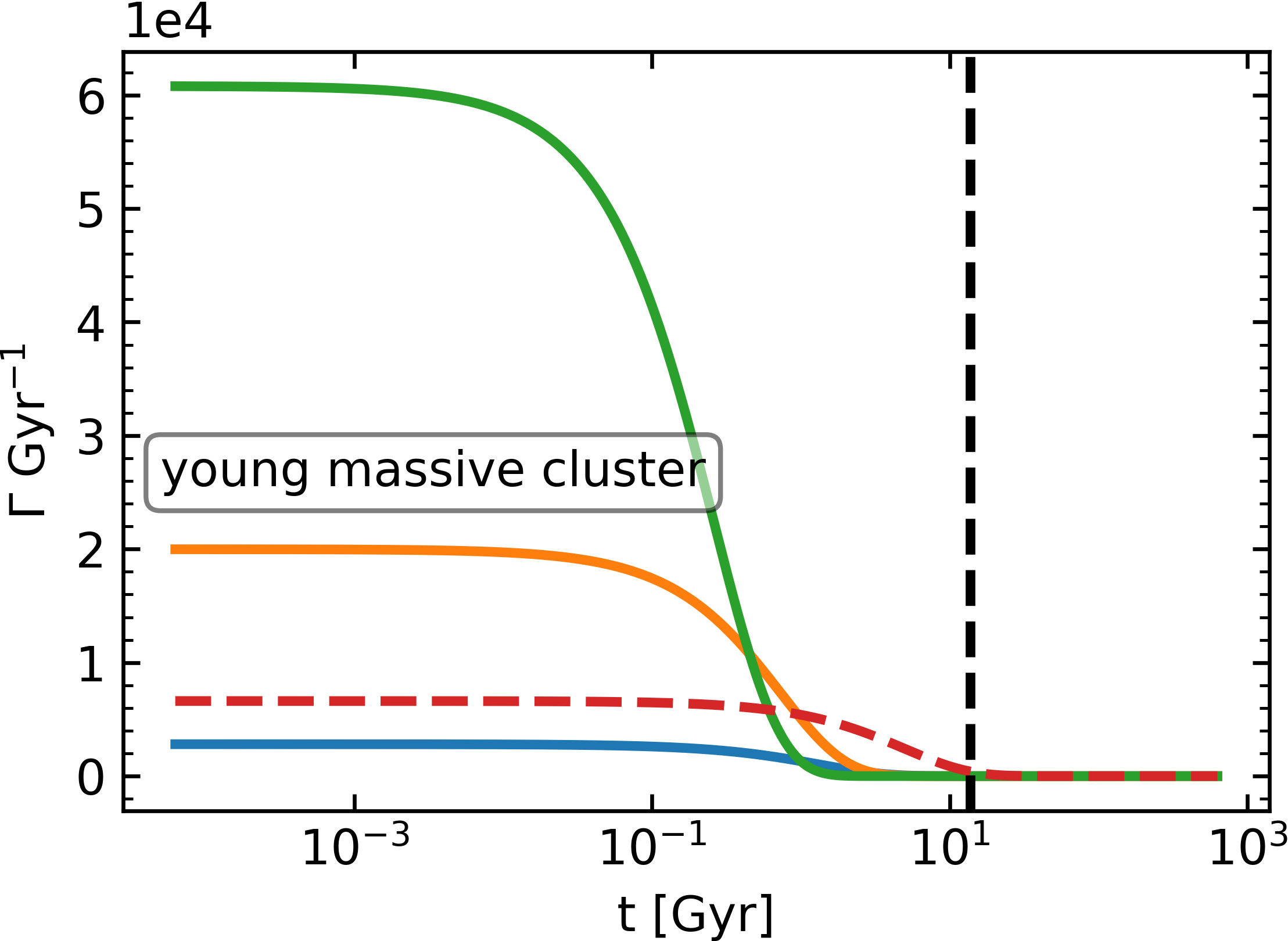}
	\includegraphics[width=0.66\columnwidth]{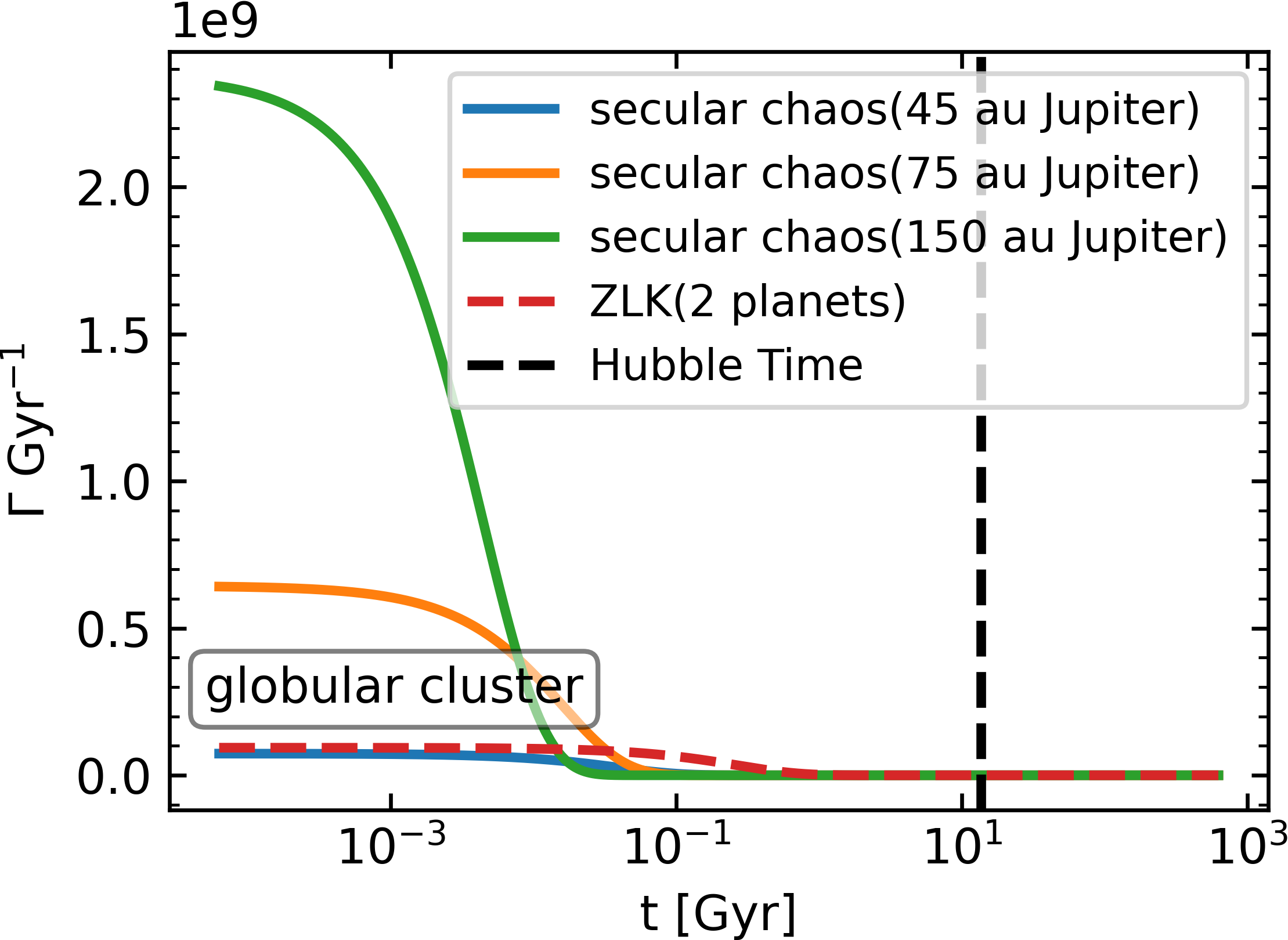}
	
    \caption{Global hot Jupiter formation rate from flyby-induced secular chaos and ZLK given by Equation~\ref{eq:grate} as a function of time. This rate is corrected for the disruption probability of 3-planet systems in dense clusters.  The \textit{left} panels show the rate for a typical open cluster with  $n=10^2$ pc$^{-3}$, $\sigma = 0.1$ km/s, $N_\star=$ 100. The \textit{middle} panels show the rate for a typical young massive cluster with  $n=10^4$ pc$^{-3}$, $\sigma = 1$ km/s, $N_\star=$ 10$^4$ while the \textit{right} panels show the rate for a typical globular cluster with  $n=10^6$ pc$^{-3}$, $\sigma = 5$ km/s, $N_\star=$ 10$^6$.  The {\em upper} panels show the rates of planetary systems with Neptune-mass outer planets, while the  {\em bottom} panels show the rates for planetary systems with Jupiter-mass outer planets.}
    \label{fig:final-rate}
\end{figure*}

Figure~\ref{fig:final-rate} shows the modified hot Jupiter formation rate, corrected for the disruption probability of a three-planet system in dense clusters. As expected, the reduction is the largest for the densest cluster (right panel of the figure).  For planetary systems with a Neptune-mass outer planet, larger systems tend to decrease $P_{\rm three}$ without significantly increasing in $\gamma$, thus leading to a lower $\Gamma$, especially in low velocity dispersion environments. For planetary systems with a Jupiter-mass outer planet, the size effect on increasing $\gamma$ is more significant, thus planetary systems of larger size tend to have larger $\Gamma$ in various velocity dispersion environments \citep[e.g.][]{Leigh11}.
We remark that this adjusted rate  assumes that the planetary system has always been part of a star cluster. However, it is possible that the system may have been part of a dense star cluster when created, but then the cluster dissolved, in which case the hot Jupiter formation rate in a three-planet system should not be corrected for the disruption probability of the three-planet system (see discussion in \citealt{Winter2020}).   

Last but importantly, we wish to put our results within	the
context	of hot Jupiter observations, which reveal that they
are in large part lonely planets, without close companions.
A uniform search with TESS data yielded no statistically valid
candidate with orbital period	$P<14$~days \citep{Hord2021}.
Previous searches had already noted the	common loneliness of hot Jupiters
out to orbital periods of $\sim 200$~days \citep{Steffen2012,Knutson2014,Endl2014}.
Our investigation of flyby-induced secular chaos has showed that {\em large} planetary sizes are
needed in order	for the	system to aquire enough	AMD to trigger high eccentricity
migration. The two outer planets, which	are at initial SMAs of 15~au and 45~au or larger,
have similar final SMAs, since planets in the secular regime do not tend to exchange orbital energy between themselves, and the tidal dissipation on the outer two planets is too weak due to their lower eccentricity.  Hence	the outer planets responsible for the hot Jupiter formation would be undetectable in transit searches.
On the other hand, radial velocity (RV) observations can detect giant planet companions out to larger distances. The RV survey by \citet{Knutson2014} estimated a total occurrence rate of $51\%\pm 10\%$ for companions with $1-13$ Jupiter masses and SMAs between 1-20~au.
Similarly, the survey by \citet{Bryan2016} estimated a total occurrence rate of companions of 
$52\% \pm 5\%$ in the mass range $1-20$ Jupiter masses and for SMAs between 5-20~au. Both these observations indicate that long-period companions to hot Jupiters are common, and may hence play a role in their formation, as expected via secular chaos.

\section{Conclusions}
In this work we have continued and expanded upon our previous study of
hot Jupiter formation in an environment with strong stellar
interactions via high-precision, $N$-body simulations.  The study by
\citet{Wang2020} found that the flyby-induced ZLK effect in two giant
planet systems can efficiently create hot Jupiters in star
clusters. In this paper, by performing high precision few-body
scatterings of stellar flybys, and following the systems with
long-term integrations inclusive of tidal dissipation and general
relativity corrections, we study the impact of stellar
perturbations on three-planet systems in star clusters. Our main
conclusions can be summarized as follows:

\begin{itemize}

\item Close stellar flybys can trigger secular chaos in multi-planet
  systems that are initially in a circular, coplanar architecture. The
  stellar flybys incline and increase the eccentricity of the planets'
  orbits and import an angular momentum deficit (AMD), which is a
  requirement for triggering secular chaos. The secular chaos after
  the flyby can drive the innermost Jupiter to an extremely eccentric
  orbit, which creates the conditions for high eccentricity tidal
  migration to form hot Jupiters.

\item
  The fraction of systems with sufficient AMD to trigger secular chaos
  after the flyby increases with both the size of the planetary system
  (namely, the semi-major axis of the third outer planet), as well as
  with the mass of the outer planet.  For a Neptune-sized outer
  planet, this fraction ranges from about 10\% to 20\% as the initial
  SMA of the planet varies between 45 and 150~au.  For the same
  range of initial SMAs, the fraction increases to the range 18\%-52\%
  if the outer planet has a Jupiter-like mass.

 \item Close stellar flybys can also trigger planet-planet scattering
  in multi-planet systems. However, only a very small fraction of
  systems have planets with sufficiently overlapping orbits to result in a
  scattering. Therefore, in multi-planet systems, flyby induced
  secular chaos is the dominant effect to trigger high eccentricity
  tidal migration.

\item For the range of outer planet SMAs and masses that we consider, 
  the formation rate of hot Jupiters per system via flyby-induced secular
    chaos in three-planet systems straddles the rate of
    flyby-induced ZLK in two-planet systems (for identical systems without
    the third outer planet). However, the formation rate via secular chaos
    increases more rapidly than the ZLK rate for denser clusters.

  \item For stellar flybys on multi-planet systems in low-velocity
  dispersion environments, each planet can be treated individually and
  their ejection rate and post-flyby orbital properties are almost
  identical as a function of the dimensionless parameter $Q/a_i$, where
  $Q$ is the distance of closest approach and $a_i$ is the semi-major axis of the
  i-th planet. Based on this result, we build a simple model to
  calculate the survivability of multi-planet systems in star clusters
  and the configurations of planetary systems after the flyby.
  As expected, multi-planet systems are preferentially populated in lower-number
  density clusters like open clusters, since the frequent
  interactions in very dense clusters tend to eject the outer planet
  of the  system in a short time. Therefore, if multi-planet dynamics
  contribute to the hot Jupiter formation  rate in star clusters, it
  is more likely in open clusters.
  
  \item{}
  Given the higher efficiency of secular chaos for larger planetary sizes and more massive outer planets, hot Jupiters formed via this mechanism are expected to be accompanied by outer giants at large orbital separations, which should be searched with radial velocity surveys.

\end{itemize}
\section*{Acknowledgements}
NWCL gratefully acknowledges the generous support of a Fondecyt Iniciaci\'on grant 11180005, in addition to financial support from Millenium Nucleus NCN19-058 (TITANs).

%%%%%%%%%%%%%%%%%%%%%%%%%%%%%%%%%%%%%%%%%%%%%%%%%%
\section*{Data Availability}
Data generated for this study can be accessed at \url{https://github.com/YihanWangAstro/Nepture}. The code used for the simulations is publicly available at \url{https://github.com/YihanWangAstro/SpaceHub}.

%%%%%%%%%%%%%%%%%%%% REFERENCES %%%%%%%%%%%%%%%%%%

% The best way to enter references is to use BibTeX:

\bibliographystyle{mnras}
\bibliography{example} % if your bibtex file is called example.bib

%%%%%%%%%%%%%%%%%%%%%%%%%%%%%%%%%%%%%%%%%%%%%%%%%%

%%%%%%%%%%%%%%%%% APPENDICES %%%%%%%%%%%%%%%%%%%%%

\appendix

\section{Some extra material}

If you want to present additional material which would interrupt the flow of the main paper,
it can be placed in an Appendix which appears after the list of references.

%%%%%%%%%%%%%%%%%%%%%%%%%%%%%%%%%%%%%%%%%%%%%%%%%%

% Don't change these lines
\bsp	% typesetting comment
\label{lastpage}
\end{document}